\documentclass[10pt,a4paper]{article}
\usepackage[utf8]{inputenc}
\usepackage[english]{babel}

\usepackage{amsmath}
\usepackage{amsfonts}
\usepackage{amssymb}

\usepackage[colorlinks,citecolor=blue,urlcolor=blue,linkcolor=blue]{hyperref}
\usepackage[left=2cm,right=2cm,top=2cm,bottom=2cm]{geometry}

\usepackage{authblk}

\usepackage{feynmp}
\newcommand{\pd}{\partial}
\newcommand{\gs}{\mathit{g}_\text{s}}
\newcommand{\korder}[1]{\mathcal{O}\left(\kappa^#1\right)}
\newcommand{\abs}[1]{\left\lvert #1 \right\rvert}

\title{FeynGrav 2.0}
\author[1,2]{B. Latosh \thanks{latosh.boris@ibs.re.kr}}
\affil[1]{Center for Theoretical Physics of the Universe, IBS, 34126 Daejeon, South Korea}
\affil[2]{Bogoliubov Laboratory of Theoretical Physics, JINR, Dubna 141980, Russia}
\date{CTPU-PTC-23-06}

\begin{document}

\maketitle

\begin{abstract}
  We present a new version of FeynGrav. The present version supports Feynman rules for matter with non-vanishing mass and $SU(N)$ Yang-Mills model. We revisit the gauge fixing procedure for gravity and derive interaction rules valid for an arbitrary gauge fixing parameter. We provide a few simple examples of calculations to illustrate package usage.
\end{abstract}

\section{Introduction}

This paper presents the recent development of the package ``FeynGrav'' \cite{Latosh:2022ydd}. The package provides a tool to operate with Feynman rules for perturbative quantum gravity within FeynCalc \cite{Mertig:1990an,Shtabovenko:2016sxi,Shtabovenko:2020gxv}. In \cite{Latosh:2022ydd} the author proposed a novel analytic approach to the derivation of Feynman rules. It provides a way to construct the Feynman rules for a wide class of gravity models. It was applied to models without supersymmetry and non-minimal coupling to gravity. Interaction rules for the massless matter of spin $0$, $1/2$, and $1$ were derived and their implementation within FeynGrav was discussed.

In this paper, we present a further development of the analytic approach proposed earlier and its implementation for FeynGrav. Firstly, we consider a matter of spin $0$, $1/2$, and $1$ with non-vanishing masses and minimal coupling to gravity. We pay particular attention to the case of a  massless vector field and revisit the issue of gauge fixing. We demonstrate that the corresponding Faddeev-Popov ghosts interact with gravitational degrees of freedom. In addition, we derive the interaction rules for scalar field potential.

Secondly, we consider the gravitational coupling to $SU(N)$ Yang-Mills model. We derive the corresponding Feynman rules and show that, similarly to the case of a single massless vector field, the Faddeev-Popov ghosts interact with the gravitational degrees of freedom. This generalization allows for the calculation of scattering amplitudes in gravity coupled to gauge fields and opens new perspectives for phenomenological investigations.

Finally, we revisit the gauge fixing procedure for gravity and introduce more general gauge fixing conditions. The corresponding gauge fixing parameter is made explicit in all calculations. In full analogy with the previous cases, the corresponding Faddeev-Popov ghosts interact with the gravitational degrees of freedom.

All models discussed in this paper are implemented within the new version of FeynGrav. Its usage is illustrated in a few physically relevant examples.

It shall be noted that there are different approaches to the derivation of the Feynman rules for gravity. For instance, in classical papers \cite{DeWitt:1967yk,DeWitt:1967ub,DeWitt:1967uc} interaction rules for three and four-graviton vertices were derived directly from the Hilbert action. A similar approach based on the Hilbert action perturbative expansion was constructed in \cite{Prinz:2020nru}. Widely-known package xAct \cite{xActBundle,Portugal:1998qi,MARTINGARCIA2008597,Brizuela:2008ra,Pitrou:2013hga,Nutma:2013zea} also provides a tool to operate with perturbative expansion within gravity models, but its applicability is mostly limited to the classical domain. We discuss opportunities to implement it within FeynGrav in the previous paper \cite{Latosh:2022ydd}. Lastly, recently another package providing a tool to operate with Feynman rules for gravity-matter coupling was created \cite{SevillanoMunoz:2022tfb}. A more detailed discussion of computer algebra application for gravity research lies beyond the scope of this paper and can be found in the following reviews \cite{MacCallum:2018csx,HEPSoftwareFoundation:2020daq}.

In this paper, we present a comprehensive study of Feynman's rules for perturbative quantum gravity, covering matter with spin $0$, $1/2$, and $1$, $SU(N)$ Yang-Mills model, and the gauge fixing procedure. In Section \ref{section_review}, we provide an overview of our approach to derive these Feynman rules, including the notations used throughout the paper. The Feynman rules for matter fields are derived and presented. In Section \ref{section_SUNYM}, we extend our analysis to $SU(N)$ Yang-Mills model coupled to gravity. We revisit the gauge fixing procedure for gravity in Section \ref{section_gravity_ghosts} and discuss the interaction of Faddeev-Popov ghosts with gravitational degrees of freedom. In Section \ref{section_FG2}, we introduce the new version of FeynGrav, which implements all the models studied in this paper, and we illustrate its usage through a few physically relevant examples. Finally, we conclude with a discussion of the prospects and further development of FeynGrav in Section \ref{section_conclusions}.

\section{Perturbative Quantum Gravity}\label{section_review}

Perturbative quantum gravity associates gravitational phenomena with small metric perturbations propagating about the flat background. In that case the complete spacetime metric $g_{\mu\nu}$ is given as the following finite expansion:
\begin{align}\label{the_perturbative_expansion}
  g_{\mu\nu} = \eta_{\mu\nu} + \kappa \, h_{\mu\nu} .
\end{align}
Here $\eta_{\mu\nu}$ is the flat metric, $h_{\mu\nu}$ are small metric perturbations with the canonical mass dimension, and $\kappa$ is the gravity coupling related with the Newton's constant $G_\text{N}$:
\begin{align}\label{the_gravitational_coupling_definition}
  \kappa^2 \overset{\text{def}}{=} 32\,\pi\, G_\text{N} .
\end{align}
Although \eqref{the_perturbative_expansion} is a finite expression, it spawns infinite perturbative expansions for the inverse metric
\begin{align}
  g^{\mu\nu} = \eta^{\mu\nu} - \kappa\, h^{\mu\nu} + \kappa^2\, h^{\mu\sigma} h_\sigma{}^\nu + \korder{3};
\end{align}
for the volume factor
\begin{align}
  \sqrt{-g}  = 1 + \cfrac{\kappa}{2} \, h -\cfrac{\kappa^2}{4} \left(h_{\mu\nu}^2 - \cfrac12\, h^2\right) + \korder{3};
\end{align}
for the Christoffel symbols
\begin{align}
  \Gamma^\alpha_{\mu\nu} = g^{\alpha\beta} \,\Gamma_{\beta\mu\nu} =\Big( \eta^{\alpha\beta} - \kappa\,h^{\alpha\beta}+\kappa^2\, h^{\alpha\sigma} h_\sigma{}^\beta +\korder{3}\Big)\,\cfrac{\kappa}{2}\,\left[ \pd_\mu h_{\nu\beta} + \pd_\nu h_{\mu\beta} - \pd_\beta h_{\mu\nu} \right];
\end{align}
and, ultimately, for the Hilbert action
\begin{align}
  \begin{split}
    \mathcal{A}_\text{H}[g_{\mu\nu}] \overset{\text{def}}{=}&\int d^4 x \sqrt{-g} \left[-\cfrac{2}{\kappa^2} \, R\right] =\mathcal{A}_\text{H}[\eta] +  \cfrac{\delta\mathcal{A}_H}{\delta g_{\mu\nu}}[\eta]\, \kappa h_{\mu\nu} + \cfrac{\delta^2 \mathcal{A}_H}{\delta g_{\mu\nu} \,\delta g_{\alpha\beta}} [\eta] ~\kappa^2 \, h_{\mu\nu} \,h_{\alpha\beta} + \korder{3} \\
    =& h_{\mu\nu} \,\mathcal{D}^{\mu\nu\alpha\beta} \square \,h_{\alpha\beta} + \kappa\, \left(\mathfrak{V}^{(3)}\right)^{\mu_1\nu_1\mu_2\nu_2\mu_3\nu_3} \,h_{\mu_1\nu_1} h_{\mu_2\nu_2} h_{\mu_3\nu_3}  + \korder{2} .
  \end{split}
\end{align}
In this formula, the Hilbert action evaluated at the flat metric vanishes. The term linear in perturbations also vanishes because the flat background delivers a minimum to the Hilbert action. The term quadratic in perturbations describes the propagation of such perturbations. All other terms of higher orders in perturbations describe their interactions.

Perturbative quantum gravity is a quantum theory of small metric perturbations $h_{\mu\nu}$ constructed with the functional integral technique. For the sake of briefness, we call quanta of the field $h_{\mu\nu}$ gravitons. Their quantum behavior is described by the following generating functional:
\begin{align}
  \begin{split}
    \mathcal{Z} \overset{\text{def}}{=}& \int\mathcal{D} [g] \, \exp\Big[i\, \mathcal{A}_\text{H}[g]\Big] \\
    =& \int \mathcal{D}[h]\, \exp\Bigg[ i\, h_{\mu\nu} \,\mathcal{D} ^{\mu\nu\alpha\beta} \square \,h_{\alpha\beta} +i\, \kappa\, \left(\mathfrak{V}^{(3)}\right)^{\mu_1\nu_1\mu_2\nu_2\mu_3\nu_3} \,h_{\mu_1\nu_1} h_{\mu_2\nu_2} h_{\mu_3\nu_3}  + \korder{2} \Bigg] .
  \end{split}
\end{align}
We shall note that this expression shall not be used directly before the gauge fixing procedure is performed. We discuss it in detail in Section \ref{section_gravity_ghosts}.

The perturbative structures of the inverse metric $g^{\mu\nu}$, the volume factor $\sqrt{-g}$, and the vierbein $\mathfrak{e}_m{}^\mu$ are described by families of $\mathcal{I}$ and $\mathcal{C}$ tensors defined in the original paper \cite{Latosh:2022ydd}. These tensors can be generated within a computer algebra system and offer a straightforward way to handle the corresponding perturbative expansions.  While their discussion is beyond the scope of this paper, they are covered in great detail in \cite{Latosh:2022ydd}.

We introduce the following notations for perturbative expansions. If a quantity $X$ is expanded in a perturbative series with respect to $\kappa\, h_{\mu\nu}$, we note the corresponding series as follows:
\begin{align}
  X = \sum\limits_{n=0}^\infty \,\kappa^n\,(X)^{\rho_1\sigma_1\cdots\rho_n\sigma_n} \, h_{\rho_1\sigma_1}\cdots h_{\rho_n\sigma_n} .
\end{align}
Here $(X)^{\rho_1\sigma_1\cdots\rho_n\sigma_n}$ notes an expression that specifies the tensor structure of a given term. To put it otherwise, it shows how indices of metric perturbations shall be contracted. In these notations perturbative expansions for $g^{\mu\nu}$ and $\sqrt{-g}$ are written as follows:
\begin{align}
  \begin{split}
    g^{\mu\nu} =& \sum\limits_{n=0}^\infty \, \kappa^n \,\left(g^{\mu\nu}\right)^{\rho_1\sigma_1\cdots\rho_n\sigma_n} \, h_{\rho_1\sigma_1}\cdots h_{\rho_n\sigma_n} ,\\
    \sqrt{-g} =& \sum\limits_{n=0}^\infty \, \kappa^n \, \left(\sqrt{-g} \right)^{\rho_1\sigma_1\cdots\rho_n\sigma_n} \, h_{\rho_1\sigma_1}\cdots h_{\rho_n\sigma_n} .
  \end{split}
\end{align}
For the sake of illustration, we present a few terms from these expressions:
\begin{align}
  \left( g^{\mu\nu} \right) &= \eta^{\mu\nu} ,&  \left(g^{\mu\nu}\right)^{\alpha\beta} &= \cfrac12\,\left(\eta^{\mu\alpha}\eta^{\nu\beta} + \eta^{\mu\beta}\eta^{\nu\alpha}\right), &   \\
  \left(\sqrt{-g} \right) &= 1 ,&  \left(\sqrt{-g}\right)^{\mu\nu} &= \cfrac12\,\eta^{\mu\nu}, \hspace{20pt} \left(\sqrt{-g}\right)^{\mu\nu\alpha\beta} =\cfrac{1}{8} \left(-\eta^{\alpha\nu} \eta^{\beta\mu}-\eta^{\alpha\mu} \eta^{\beta\nu}+\eta^{\alpha\beta} \eta^{\mu\nu}\right).\nonumber
\end{align}

All of the interaction rules presented in this paper have been derived using perturbative techniques as described above. It is worth noting that this approach can be extended to supersymmetric models and models with non-minimal gravitational coupling. These cases will be discussed in future works.

Let us briefly review the construction of the Feynman rules for a single scalar field, a single Dirac fermion, and a single vector field. The scalar and Dirac field cases were covered in detail in the original paper, so we will only briefly touch upon them. The construction of Feynman rules for a vector field is more intricate due to the gauge fixing and will be discussed in more depth

\subsection{Single scalar field}

A single free scalar field minimally coupled to gravity is described by the following action:
\begin{align}
  \begin{split}
    \mathcal{A}_{s=0} =& \int d^4 x \sqrt{-g} \left[ \cfrac12\,\left(\nabla\phi\right)^2 - \cfrac{m_\text{s}^2}{2} \, \phi^2 \right] = \int d^4 x \left[ \cfrac12\, \sqrt{-g}\, g^{\mu\nu} \, \pd_\mu\phi \, \pd_\nu\phi - \cfrac{m^2_\text{s}}{2}\,\sqrt{-g} \, \phi^2 \right].
  \end{split}
\end{align}
Here $m_\text{s}$ is the scalar field mass. Its perturbative expansion in the momentum representation reads:
\begin{align}
  \begin{split}
    \mathcal{A}_{s=0} =& \sum\limits_{n=0}^\infty \int \cfrac{d^4 p_1}{(2\pi)^4} \cfrac{d^4 p_2}{(2\pi)^4} \prod\limits_{i=1}^n \cfrac{d^4 k_i}{(2\pi)^4}\,(2\pi)^4 \delta\left(p_1+p_2+\sum k_i\right) h_{\rho_1\sigma_1}(k_1) \cdots h_{\rho_n\sigma_n}(k_n)\\
    &\times \kappa^n \left[ -\cfrac12\, \left(\sqrt{-g}\, g^{\mu\nu}\right)^{\rho_1\sigma_1\cdots\rho_n\sigma_n}  I_{\mu\nu\alpha\beta} (p_1)^\alpha (p_2)^\beta - \cfrac{m^2_\text{s}}{2} \left(\sqrt{-g}\right)^{\rho_1\sigma_1\cdots\rho_n\sigma_n} \right]\,\phi(p_1) \phi(p_2)\,.
  \end{split}
\end{align}
Here $k_i$ are momenta of gravitons, $p_1$ and $p_2$ are momenta of scalars, and $I$ tensor contracts indices of the metric and momenta in a symmetric way:
\begin{align}
  I^{\mu\nu\alpha\beta} = \cfrac12\, \left( \eta^{\mu\alpha}\eta^{\nu\beta} + \eta^{\mu\beta}\eta^{\nu\alpha}\right) .
\end{align}

The background contribution of this expression describes the scalar field propagator:
\begin{align}\label{scalar_propagator}
  \begin{gathered}
    \begin{fmffile}{Diag01}
      \begin{fmfgraph}(30,30)
        \fmfleft{L}
        \fmfright{R}
        \fmf{dashes}{L,R}
      \end{fmfgraph}
    \end{fmffile}
  \end{gathered}
  = i \, \cfrac{1}{p^2 - m_\text{s}^2} ~.
\end{align}
The other parts of this expression define rules for gravitons coupling to the scalar field kinetic energy:
\begin{align}
  \nonumber \\
  \begin{gathered}
    \begin{fmffile}{FR_S_1}
      \begin{fmfgraph*}(40,40)
        \fmfleft{L1,L2}
        \fmfright{R1,R2}
        \fmf{dbl_wiggly}{L1,V}
        \fmf{dbl_wiggly}{L2,V}
        \fmfdot{V}
        \fmf{dashes}{V,R1}
        \fmf{dashes}{V,R2}
        \fmffreeze
        \fmf{dots}{L1,L2}
        \fmflabel{$p_1$}{R1}
        \fmflabel{$p_2$}{R2}
        \fmflabel{$\rho_1\sigma_1$}{L1}
        \fmflabel{$\rho_n\sigma_n$}{L2}
      \end{fmfgraph*}
    \end{fmffile}
  \end{gathered}
  = -i\, \kappa^n \,\left[ \left(\sqrt{-g}\, g^{\mu\nu} \right)^{\rho_1\sigma_1\cdots\rho_n\sigma_n} \, I_{\mu\nu\alpha\beta} (p_1)^\alpha (p_2)^\beta + m_\text{s}^2\, \left( \sqrt{-g}\right)^{\rho_1\sigma_1\cdots\rho_n\sigma_n} \right]. \\ \nonumber
\end{align}
Here and further, all momenta are directed inwards and connected by conservation law. The dotted line on the left part of the diagram notes the presence of $n\geq 1$ graviton lines. This expression is symmetric with respect to the scalar field momenta. In the rest of this paper, we present expressions that are also symmetric with respect to momenta.

The gravitational coupling of a scalar field potential energy is derived similarly. The scalar field potential $V(\phi)$ shall be expanded in a power series with respect to the scalar field $\phi$. Each term of this expansion corresponds to a separate scalar field self-interaction coupled to gravity. Therefore, it is sufficient to derive the interaction rule for a single power-law potential. Let us consider the following power-law potential with $p \geq 3$ being a whole number, and $\lambda_p$ being a coupling with the mass dimension $4-p$:
\begin{align}
  \mathcal{A}_{s=0,\text{potential}} = \int d^4 x \sqrt{-g} \left[ \cfrac{\lambda_p}{p!} ~ \phi^p \right] = \int d^4 x \left[ \sqrt{-g} ~ \cfrac{\lambda_p}{p!} ~ \phi^p \right].
\end{align}
In the momentum representation, this action becomes:
\begin{align}
  \begin{split}
    \mathcal{A}_{s=0,\text{potential}} =& \sum\limits_{n=0}^\infty\int \prod\limits_{j=1}^p \cfrac{d^4 q_j}{(2\pi)^4} \prod\limits_{i=0}^n \cfrac{d^4 k_i}{(2\pi)^4} \, (2\pi)^4 \,\delta\Big( \sum q_j + \sum k_i \Big)\,h_{\rho_1\sigma_1}(k_1)\cdots h_{\rho_n\sigma_n}(k_n) \\
    &\times \kappa^n \,\cfrac{\lambda_p}{p!} \, \left(\sqrt{-g}\right)^{\rho_1\sigma_1\cdots\rho_n\sigma_n} \phi(q_1) \cdots \phi(q_p)\,.
  \end{split}
\end{align}
The corresponding interaction rule reads:
\begin{align}
  \nonumber \\
  \begin{gathered}
    \begin{fmffile}{FR_S_Power}
      \begin{fmfgraph*}(40,40)
        \fmfleft{L1,L2}
        \fmfright{R1,R2}
        \fmf{dbl_wiggly}{L1,V}
        \fmf{dbl_wiggly}{L2,V}
        \fmf{dashes}{R1,V}
        \fmf{dashes}{R2,V}
        \fmfdot{V}
        \fmffreeze
        \fmf{dots}{L1,L2}
        \fmf{dots}{R1,R2}
        \fmflabel{$\rho_1\sigma_1$}{L1}
        \fmflabel{$\rho_n\sigma_n$}{L2}
        \fmflabel{$q_1$}{R1}
        \fmflabel{$q_p$}{R2}
      \end{fmfgraph*}
    \end{fmffile}
  \end{gathered}
  = i\,\kappa^n \, \lambda_p \, \left(\sqrt{-g}\right)^{\rho_1\sigma_1\cdots\rho_n\sigma_n} . \\ \nonumber
\end{align}
This expression can be used for any whole $p \geq 3$, so it completely describes the gravitational coupling of scalar field potentials.

The healthy scalar field interactions described in this paper are just a small subset of the broader range of interactions described by the Horndeski and Beyond Horndeski models \cite{Horndeski:1974wa,Horndeski:1976gi,Kobayashi:2011nu,Zumalacarregui:2013pma,Gleyzes:2014dya,BenAchour:2016cay,Kobayashi:2019hrl}. Feynman rules for these models are beyond the scope of this paper and will be discussed in future publications.

\subsection{Single Dirac field}

A single Dirac field minimally coupled to gravity is described by the following action:
\begin{align}
  \begin{split}
    \mathcal{A}_{s=1/2} =& \int d^4 x \sqrt{-g} \left[ \overline{\psi} \left( i\, \widehat{\nabla}\right) \psi - m_\text{f}\, \overline{\psi}\,\psi \right] \\
    =& \int d^4 x \left[ \sqrt{-g} \,\mathfrak{e}_m{}^\mu \, \frac12\, \left( i\,\overline{\psi}\, \gamma^m \,\nabla_\mu \psi - i\, \nabla_\mu\overline{\psi} \,\gamma^m \,\psi\right) - m_\text{f}\, \sqrt{-g}\, \overline{\psi} \,\psi \right] .
  \end{split}
\end{align}
Here $m_\text{f}$ is the fermion mass,  $\mathfrak{e}_m{}^\mu$ is the vierbein, and $\nabla$ is the fermionic covariant derivative. We discuss the construction of spinors in a curved spacetime alongside their perturbative treatment in the previous paper \cite{Latosh:2022ydd} (see also \cite{Shapiro:2016pfm,stepanyants2009}). The following theorem specifies the perturbative structure of this action \cite{Latosh:2022ydd}:
\begin{align}
  \begin{split}
    \mathcal{A}_{s=1/2}  =& \int d^4 x \left[ \sqrt{-g} \, \mathfrak{e}_m{}^\mu \, \frac12\, \left( i\,\overline{\psi}\, \gamma^m \,\pd_\mu \psi - i\, \pd_\mu\overline{\psi} \,\gamma^m \,\psi\right) - m_\text{f}\, \sqrt{-g}\, \overline{\psi} \,\psi \right]\\
    =& \sum\limits_{n=0}^\infty \int \cfrac{d^4 p_1}{(2\pi)^4} \cfrac{d^4 p_2}{(2\pi)^4} \prod\limits_{i=0}^n \cfrac{d^4 k_i}{(2\pi)^4} \,(2\pi)^4 \delta\left(p_1+p_2+\sum k_i\right) h_{\rho_1\sigma_1}(k_1) \cdots h_{\rho_n\sigma_n} (k_n)\\
    &\times \kappa^n~ \overline{\psi}(p_2) \left[ \left(\sqrt{-g}\, \mathfrak{e}_m{}^\mu\right)^{\rho_1\sigma_1\cdots\rho_n\sigma_n}\,\frac12 \, (p_1-p_2)_\mu \gamma^m - \left(\sqrt{-g}\right)^{\rho_1\sigma_1\cdots\rho_n\sigma_n} m_\text{f} \right] \psi(p_1).
  \end{split}
\end{align}
The background part of this expansion corresponds to the fermion propagator:
\begin{align}
  \begin{gathered}
    \begin{fmffile}{Diag02}
      \begin{fmfgraph}(35,35)
        \fmfleft{L}
        \fmfright{R}
        \fmf{fermion}{L,R}
      \end{fmfgraph}
    \end{fmffile}
  \end{gathered}
  \hspace{10pt}= i ~ \cfrac{p_m \,\gamma^m +m_\text{f}}{p^2 - m_\text{f}^2} \,. 
\end{align}
The other terms describe the following interaction rules:
\begin{align}\label{rule_F_1}
  \nonumber \\
  \begin{gathered}
    \begin{fmffile}{FR_F_1}
      \begin{fmfgraph*}(40,40)
        \fmfleft{L1,L2}
        \fmfright{R1,R2}
        \fmf{dbl_wiggly}{L1,V}
        \fmf{dbl_wiggly}{L2,V}
        \fmfdot{V}
        \fmf{fermion}{R1,V}
        \fmf{fermion}{V,R2}
        \fmffreeze
        \fmf{dots}{L1,L2}
        \fmflabel{$p_1$}{R1}
        \fmflabel{$p_2$}{R2}
        \fmflabel{$\rho_1\sigma_1$}{L1}
        \fmflabel{$\rho_n\sigma_n$}{L2}
      \end{fmfgraph*}
    \end{fmffile}
  \end{gathered}
  = i\, \kappa^n \,\left[  \cfrac12\, \left(\sqrt{-g}\, \mathfrak{e}_m{}^\mu\right)^{\rho_1\sigma_1\cdots\rho_n\sigma_n} \, (p_1-p_2)_\mu \gamma^m - \left(\sqrt{-g}\right)^{\rho_1\sigma_1\cdots\rho_n\sigma_n} m_\text{f} \right]. \\ \nonumber
\end{align}
As it was noted above, on this diagram all momenta are directed inwards, so $p_1$ notes an in-going momentum of a fermion, $p_2$ notes an in-out momentum of an anti-fermion. Moreover, this expression is applicable for the $SU(N)$ Yang-Mills model considered below.

\subsection{Single vector field}

The treatment of a vector field within the quantum field theory (and perturbative quantum gravity) is sensitive to the vector field mass. A massless vector field admits the gauge symmetry, so the gauge fixing shall be performed. If a vector field has a non-vanishing mass, then the gauge symmetry is not present and gauge fixing is not required.

We start with the case of a vector field with a non-vanishing mass, also known as the Proca field. Such a field coupled with gravity is described by the following action:
\begin{align}
  \begin{split}
    \mathcal{A}_{s=1,m_\text{v}} =& \int d^4 x \sqrt{-g} \left[ -\cfrac14\, F_{\mu\nu}F^{\mu\nu} + \cfrac{m_\text{v}^2}{2} \, A_\mu \,A^\mu \right].
  \end{split}
\end{align}
Here $F_{\mu\nu} = \pd_\mu A_\nu - \pd_\nu A_\mu$ is the field tensor, $m_\text{v}$ is the vector field mass. The perturbative expansion of this action in the momentum representation reads:
\begin{align}
  \begin{split}
    \mathcal{A}_{s=1,m_\text{v}}=& \sum\limits_{n=0}^\infty \int \cfrac{d^4 p_1}{(2\pi)^4} \cfrac{d^4 p_2}{(2\pi)^4} \prod\limits_{i=0}^n \cfrac{d^4 k_i}{(2\pi)^4} \,(2\pi)^4 \delta\left(p_1+p_2+\sum k_i\right) h_{\rho_1\sigma_1}(k_1) \cdots h_{\rho_n\sigma_n} (k_n)  \\
    &\times \kappa^n \,\Bigg[\cfrac14\, \left(\sqrt{-g}\, g^{\mu\alpha}g^{\nu\beta}\right)^{\rho_1\sigma_1\cdots\rho_n\sigma_n}~(p_1)_{\mu_1} (p_2)_{\mu_2}  ~\big(F_{\mu\nu}\big)^{\mu_1\lambda_1}\big(F_{\alpha\beta}\big)^{\mu_2\lambda_2}  \\
      & \hspace{30pt}+ \cfrac{m_\text{v}^2}{2} \left( \sqrt{-g}\, g^{\lambda_1\lambda_2} \right)^{\rho_1\sigma_1\cdots\rho_n\sigma_n} \Bigg] \,A_{\lambda_1}(p_1) \, A_{\lambda_2}(p_2) .
  \end{split}
\end{align}
Here we introduced the following notations:
\begin{align}
  F_{\mu\nu} &= -i\,p_\sigma \, \big(F_{\mu\nu}\big)^{\sigma\lambda} \,A_{\lambda}(p) , & \big(F_{\mu\nu}\big)^{\sigma\lambda} &\overset{\text{def}}{=} \delta^\sigma_\mu \, \delta^\lambda_\nu - \delta^\sigma_\nu \, \delta^\lambda_\mu .
\end{align}
This expression spawns the standard Proca propagator:
\begin{align}\label{Proca_propagator}
  \begin{gathered}
    \begin{fmffile}{Diag03}
      \begin{fmfgraph*}(30,30)
        \fmfleft{L}
        \fmfright{R}
        \fmf{photon}{L,R}
        \fmflabel{$\mu$}{L}
        \fmflabel{$\nu$}{R}
      \end{fmfgraph*}
    \end{fmffile}
  \end{gathered}
  \hspace{13pt}=(-i)\,\cfrac{ ~ \eta_{\mu\nu} - \cfrac{p_\mu\,p_\nu}{m_\text{v}^2} ~ }{p^2 - m_\text{v}^2}\,.
\end{align}
The interaction rules describing gravitons coupling to the Proca field kinetic energy is given by the following expression:
\begin{align}\label{rule_V_1}
  \nonumber \\
  \begin{gathered}
    \begin{fmffile}{FR_V_1}
      \begin{fmfgraph*}(40,40)
        \fmfleft{L1,L2}
        \fmfright{R1,R2}
        \fmf{dbl_wiggly}{L1,V}
        \fmf{dbl_wiggly}{L2,V}
        \fmfdot{V}
        \fmf{photon}{R1,V}
        \fmf{photon}{V,R2}
        \fmffreeze
        \fmf{dots}{L1,L2}
        \fmflabel{$p_1,\lambda_1,m_\text{v}$}{R1}
        \fmflabel{$p_2,\lambda_2,m_\text{v}$}{R2}
        \fmflabel{$\rho_1\sigma_1$}{L1}
        \fmflabel{$\rho_n\sigma_n$}{L2}
      \end{fmfgraph*}
    \end{fmffile}
  \end{gathered}
  \begin{split}
    \hspace{50pt}=  i\, \kappa^n \Bigg[ & \cfrac12\, \left(\sqrt{-g}\, g^{\mu\alpha}g^{\nu\beta}\right)^{\rho_1\sigma_1\cdots\rho_n\sigma_n} (p_1)_{\mu_1} (p_2)_{\mu_2} \big(F_{\mu\nu}\big)^{\mu_1\lambda_1} \big(F_{\alpha\beta}\big)^{\mu_2\lambda_2} \\
      & + m_\text{v}^2 \left( \sqrt{-g}\, g_{\lambda_1\lambda_2} \right)^{\rho_1\sigma_1\cdots\rho_n\sigma_n} \Bigg].
  \end{split}\\ \nonumber
\end{align}

To proceed with the massless case we shall briefly recall the Faddeev-Popov prescription for gauge theories \cite{Faddeev:1967fc,Peskin:1995ev,Weinberg:1995mt,Weinberg:1996kr,Weinberg:2000cr}. A quantum vector field is described by the following generating functional:
\begin{align}
  \mathcal{Z} = \int \mathcal{D}[A] \exp\Big[i\,\mathcal{A}[A] \Big].
\end{align}
Here the integration is performed over all conceivable fields. The normalization factor is omitted for the sake of simplicity. Firstly, one adds a new term to the microscopic action:
\begin{align}
  \mathcal{Z} = \int \mathcal{D}[A]  \exp\Big[i\,\mathcal{A}[A] \Big] \int\mathcal{D}[\omega] \exp\left[\cfrac{i}{2}\,\epsilon\,\omega^2 \right] = \int\mathcal{D}[A]\mathcal{D}[\omega] \exp\left[ i\, \mathcal{A} +\cfrac{i}{2} \, \epsilon\,\omega^2\right] .
\end{align}
Here $\omega$ is an arbitrary scalar, $\epsilon$ is a free gauge fixing parameter. The new contribution is a Gauss-like integral so its introduction merely changes the (omitted) normalization factor.

Secondly, one splits the integration volume:
\begin{align}
  \int \mathcal{D}[A] =\int \mathcal{D}[\zeta] \int \mathcal{D}[\mathbb{A}] \delta\left( \mathcal{G} - \omega \right) \det\Delta\,.
\end{align}
Here $\mathcal{G}$ is the gauge fixing condition; the new field variable $\mathbb{A}$, the gauge transformation parameter $\zeta$, and the field variable $A$ are related as follows:
\begin{align}
  A_\mu = \mathbb{A}_\mu + \pd_\mu \zeta .
\end{align}
The integration over $\mathbb{A}$ is performed over all conceivable fields, but because of the $\delta$ function from each class of physically equivalent potentials only a single representative contributes to the integral. Therefore, the integration over $\mathbb{A}$ accounts not for all conceivable potential, but for all conceivable configurations of physical fields. The last term $\det\Delta$ is the Faddeev-Popov determinant which preserves the invariance of the integration measure. The corresponding differential operator $\Delta$ is defined as follows:
\begin{align}
  \Delta \overset{\text{def}}{=} \cfrac{\delta\mathcal{G}}{\delta \zeta}\,.
\end{align}

Finally, one performs integrations and obtains the following expression for the generating functional:
\begin{align}
  \begin{split}
    \mathcal{Z} &= \int \mathcal{D}[\mathbb{A}]\mathcal{D}[\omega]\mathcal{D}[\zeta] \left(\det\Delta\right) ~ \delta\left(\mathcal{G} - \omega \right) \exp\left[ i\, \mathcal{A} +\cfrac{i}{2} \, \epsilon\,\omega^2\right] \\
    &= \int \mathcal{D}[\mathbb{A}] \left(\det\Delta\right) \exp\left[ i\, \mathcal{A} +\cfrac{i}{2} \, \epsilon\,\mathcal{G}^2 \right] \\
    &=\int\mathcal{D}[c]\mathcal{D}[\overline{c}]\mathcal{D}[\mathbb{A}] \exp\left[ i \,\overline{c} \, \Delta \, c + i \, \mathcal{A} + \cfrac{i}{2}\,\epsilon \, \mathcal{G}^2 \right] .
  \end{split}
\end{align}
Here $\overline{c}$, $c$ are scalar anticommuting Faddeev-Popov ghosts that are introduced to account for the Faddeev-Popov determinant. The integration over the gauge parameter $\zeta$ is included in the normalization factor and omitted. This prescription produces a generating functional suitable for a consistent treatment of gauge models. 

We use the standard Lorentz gauge fixing condition for the sake of simplicity. In a curved spacetime, it becomes:
\begin{align}
  g^{\mu\nu}\,\nabla_\mu A_\nu =0 \leftrightarrow g^{\mu\nu}\,\pd_\mu A_\nu - g^{\mu\nu}\, \Gamma^\sigma_{\mu\nu} A_\sigma = 0 .
\end{align}
The gauge invariant part of the action admits the following perturbative expansion in the momentum representation:
\begin{align}
  \begin{split}
    \mathcal{A}_{s=1,m_\text{v}=0} =& \int d^4 x \sqrt{-g}\left[ -\cfrac14\,g^{\mu\alpha} g^{\nu\beta}\,F_{\mu\nu}F_{\alpha\beta}\right] \\
    =& \sum\limits_{n=0}^\infty\int\cfrac{d^4p_1}{(2\pi)^4}\cfrac{d^4p_2}{(2\pi)^4}\prod\limits_{i=1}^n \cfrac{d^4k_i}{(2\pi)^4} \,(2\pi)^4 \,\delta\big(p_1+p_2+\sum k_i \big) h_{\rho_1\sigma_1}(k_1)\cdots h_{\rho_n\sigma_n}(k_n)\\
    & \times \,\kappa^n \, \left[ \cfrac14\,\left(\sqrt{-g} \,g^{\mu\alpha}g^{\nu\beta}\right)^{\rho_1\sigma_1\cdots\rho_n\sigma_n}\,(p_1)_{\mu_1}(p_2)_{\mu_2}\left(F_{\mu\nu}\right)^{\mu_1\lambda_1} \left(F_{\alpha\beta}\right)^{\mu_2\lambda_2}\right] A_{\lambda_1}(p_1) A_{\lambda_2}(p_2).
  \end{split}
\end{align}
This expression matches the expression for the Proca field with $m_\text{v}=0$. The gauge fixing term naturally splits into three terms:
\begin{align}\label{gauge_fixing_vector}
  \begin{split}
    \mathcal{A}_\text{gf} =& \int d^4 x \sqrt{-g}\left[ \cfrac{\epsilon}{2}\,\nabla_{\lambda_1} A^{\lambda_1} \, \nabla_{\lambda_2} A^{\lambda_2} \right]\\
    =& \cfrac{\epsilon}{2} \int d^4 x \left(\sqrt{-g}\,g^{\sigma_1\lambda_1}g^{\sigma_2\lambda_2}\right)\,\pd_{\sigma_1} A_{\lambda_1} \, \pd_{\sigma_2} A_{\lambda_2} -\epsilon \int d^4 x \left(\sqrt{-g}\,g^{\mu\nu}g^{\sigma_1\lambda_1}g^{\sigma_2\lambda_2}\right) \,\Gamma_{\sigma_1\mu\nu} \, A_{\lambda_1} \pd_{\sigma_2} A_{\lambda_2}\\
    &+\cfrac{\epsilon}{2}\int d^4 x \left( \sqrt{-g}\, g^{\mu\nu} g^{\alpha\beta} g^{\sigma_1\lambda_1} g^{\sigma_2\lambda_2} \right)\,\Gamma_{\sigma_1\mu\nu} \Gamma_{\sigma_2\alpha\beta} \,A_{\lambda_1} A_{\lambda_2}.
  \end{split}
\end{align}
Here we use the standard definition of the Christoffel symbols with only lower indices
\begin{align}
  \Gamma_{\mu\alpha\beta} \overset{\text{def}}{=} g_{\mu\nu} \,\Gamma^\nu_{\alpha\beta} = \cfrac12\,\left(\pd_\alpha g_{\beta\mu} + \pd_\beta g_{\alpha\mu} - \pd_\mu g_{\alpha\beta}\right).
\end{align}
In contrast with $\Gamma^\mu_{\alpha\beta}$ these symbols admit a finite perturbative expansion:
\begin{align}
  \begin{split}
    \Gamma_{\mu\alpha\beta} =& \cfrac{\kappa}{2} \left[ \pd_\alpha h_{\beta\mu} + \pd_\beta h_{\alpha\mu} - \pd_\mu h_{\alpha\beta}\right] \Leftrightarrow \kappa \, (-i)\,p_\lambda \left(\Gamma_{\mu\alpha\beta}\right)^{\lambda\rho\sigma} h_{\rho\sigma}(p) \,, \\
    \left(\Gamma_{\mu\alpha\beta}\right)^{\lambda\rho\sigma} =& \cfrac12\left[ \delta^\lambda_\alpha I_{\beta\mu}{}^{\rho\sigma} + \delta^\lambda_\beta I_{\alpha\mu}{}^{\rho\sigma} - \delta^\lambda_\mu I_{\alpha\beta}{}^{\rho\sigma} \right].
  \end{split}
\end{align}
In the momentum representation the gauge fixing term reads:
\begin{align}
  \begin{split}
    \mathcal{A}_\text{gf} =&  \sum\limits_{n=0}^\infty\int\cfrac{d^4 p_1}{(2\pi)^4}\cfrac{d^4p_2}{(2\pi)^4}\prod\limits_{i=1}^n \cfrac{d^4k_i}{(2\pi)^4}\,(2\pi)^4\delta \big(p_1+p_2+\sum k_i \big) \,h_{\rho_1\sigma_1}(k_1) \cdots h_{\rho_n\sigma_n}(k_n) A_{\lambda_1}(p_1) A_{\lambda_2}(p_2)\\
    &\times \,\kappa^n\,\left(\sqrt{-g}\, g^{\mu_1\lambda_1} g^{\mu_2\lambda_2}\right)^{\rho_1\sigma_1\cdots\rho_n\sigma_n} \left[ -\cfrac{\epsilon}{2}\, (p_1)_{\mu_1} (p_2)_{\mu_2}\right] \\
    +&\sum\limits_{n=1}^\infty\int\cfrac{d^4 p_1}{(2\pi)^4}\cfrac{d^4p_2}{(2\pi)^4}\prod\limits_{i=1}^n \cfrac{d^4k_i}{(2\pi)^4}\,(2\pi)^4\delta \big(p_1+p_2+\sum k_i \big) \,h_{\rho_1\sigma_1}(k_1) \cdots h_{\rho_n\sigma_n}(k_n) A_{\lambda_1}(p_1) A_{\lambda_2}(p_2)\\
    & \times\,\kappa^n\,\left( \sqrt{-g}\,g^{\mu\nu} g^{\mu_1\lambda_1} g^{\mu_2\lambda_2} \right)^{\rho_2\sigma_2\cdots\rho_n\sigma_n}\Big[ \epsilon \, \left(\Gamma_{\mu_1\mu\nu}\right)^{\sigma\rho_1\sigma_1}\,(k_1)_\sigma \, (p_2)_{\mu_2}\Big] \\
    +& \sum\limits_{n=2}^\infty\int\cfrac{d^4 p_1}{(2\pi)^4}\cfrac{d^4p_2}{(2\pi)^4}\prod\limits_{i=1}^n \cfrac{d^4k_i}{(2\pi)^4}\,(2\pi)^4\delta \big(p_1+p_2+\sum k_i \big) \,h_{\rho_1\sigma_1}(k_1) \cdots h_{\rho_n\sigma_n}(k_n) \,A_{\lambda_1}(p_1) A_{\lambda_2}(p_2)\\
    &\times\,\kappa^n\,\left(\sqrt{-g}\,g^{\mu\nu} g^{\alpha\beta} g^{\mu_1\lambda_1} g^{\mu_2\lambda_2} \right)^{\rho_3\sigma_3\cdots\rho_n\sigma_n} \left[ - \cfrac{\epsilon}{2}\, (k_1)_{\tau_1}(k_2)_{\tau_2} \,\big(\Gamma_{\mu_1\mu\nu} \big)^{\tau_1\rho_1\sigma_1} \big( \Gamma_{\mu_2\alpha\beta}\big)^{\tau_2\rho_2\sigma_2}\right].
  \end{split}
\end{align}

In full analogy with the previous cases, the background part of this expression corresponds to the following propagator\footnote{It matches the expression for the vector propagator given in FeynCalc with $\epsilon_\text{FeynCalc} = -1/\epsilon_\text{FeynGrav} $}:
\begin{align}\label{Maxwell_propagator}
  \begin{gathered}
    \begin{fmffile}{Diag04}
      \begin{fmfgraph*}(30,30)
        \fmfleft{L}
        \fmfright{R}
        \fmf{photon}{L,R}
        \fmflabel{$\mu$}{L}
        \fmflabel{$\nu$}{R}
      \end{fmfgraph*}
    \end{fmffile}
  \end{gathered}
  \hspace{20pt} = i ~ \cfrac{ -\eta_{\mu\nu} + \left(1+ \cfrac{1}{\epsilon}\right) \cfrac{p_\mu \, p_\nu}{p^2} }{p^2} ~.
\end{align}
The interaction rules are given by the following formula
\begin{align*}
  \nonumber \\
  \begin{gathered}
    \begin{fmffile}{FR_V_2}
      \begin{fmfgraph*}(40,40)
        \fmfleft{L1,L2}
        \fmfright{R1,R2}
        \fmf{dbl_wiggly}{L1,V}
        \fmf{dbl_wiggly}{L2,V}
        \fmfdot{V}
        \fmf{photon}{R1,V}
        \fmf{photon}{V,R2}
        \fmffreeze
        \fmf{dots}{L1,L2}
        \fmflabel{$p_1,\lambda_1$}{R1}
        \fmflabel{$p_2,\lambda_2$}{R2}
        \fmflabel{$\rho_1\sigma_1,k_1$}{L1}
        \fmflabel{$\rho_n\sigma_n,k_n$}{L2}
      \end{fmfgraph*}
    \end{fmffile}
  \end{gathered}
\end{align*}
\begin{align}\label{rule_V_2}
  \begin{split}
    = i\, \kappa^n \Bigg[& \cfrac12\,\left(\sqrt{-g}\,g^{\mu\alpha}g^{\nu\beta}\right)^{\rho_1\sigma_1\cdots\rho_n\sigma_n}\,(p_1)_{\sigma_1}(p_2)_{\sigma_2} \big(F_{\mu\nu}\big)^{\sigma_1\lambda_1} \big(F_{\alpha\beta}\big)^{\sigma_2\lambda_2} \\
      & -\epsilon \, \left(\sqrt{-g} \, g^{\mu_1\lambda_1} g^{\mu_2\lambda_2}\right)^{\rho_1\sigma_1\cdots\rho_n\sigma_n} (p_1)_{\mu_1} (p_2)_{\mu_2}  \\
      & +\epsilon\left\{ \left(\sqrt{-g}\,g^{\mu\nu}g^{\mu_1\lambda_1}g^{\mu_2\lambda_2}\right)^{\rho_2\sigma_2\cdots\rho_n\sigma_n}\,(k_1)_\sigma \left[ (p_2)_{\mu_2}\big(\Gamma_{\mu_1\mu\nu}\big)^{\sigma\rho_1\sigma_1} + (p_1)_{\mu_1}\big(\Gamma_{\mu_2\mu\nu}\big)^{\sigma\rho_1\sigma_1} \right] + \cdots \right\}\\
      & -\cfrac{\epsilon}{2} \Bigg\{ \left(\sqrt{-g}\,g^{\mu\nu}g^{\alpha\beta} g^{\mu_1\lambda_1} g^{\mu_2\lambda_2}\right)^{\rho_3\sigma_3\cdots\rho_n\sigma_n} \left[ (k_1)_{\tau_1}\,(k_2)_{\tau_2} \big(\Gamma_{\mu_1\mu\nu}\big)^{\tau_1\rho_1\sigma_1} \big(\Gamma_{\mu_2\alpha\beta}\big)^{\tau_2\rho_2\sigma_2} \right.\\
        & \hspace{190pt} \left.+ (k_1)_{\tau_2}\,(k_2)_{\tau_1} \big(\Gamma_{\mu_2\mu\nu}\big)^{\tau_1\rho_2\sigma_2} \big(\Gamma_{\mu_1\alpha\beta}\big)^{\tau_2\rho_1\sigma_1}   \right] + \cdots \Bigg\}\Bigg].
    \end{split}
\end{align}
In this expression the dots note terms that make the expression symmetric with respect to graviton momenta. The last term contributes only to vertices with $n\geq 2$ gravitons.

The ghost sector of the theory shall be treated as follows. The Faddeev-Popov differential operator $\Delta$ reduces to the D'Alamber operator in curved spacetime:
\begin{align}
  \Delta = \cfrac{\delta}{\delta \zeta} ~ \nabla_\mu\left(A^\mu + \nabla^\mu \zeta\right) = g^{\mu\nu}\nabla_\mu\nabla_\nu\,.
\end{align}
Therefore, the ghost part of the generating functional describes a single massless scalar ghost coupled to gravity:
\begin{align}
  \begin{split}
    \mathcal{Z}_\text{ghost} =& \int\mathcal{D}[c]\mathcal{D}[\overline{c}] \exp\left[i \,\int d^4 x \sqrt{-g} \, \left( \overline{c}\, \square\, c  \right)\right] \\
    =&  \int\mathcal{D}[c]\mathcal{D}[\overline{c}] \, \exp\left[ - i\, \int d^4 x \, \sqrt{-g}\, g^{\mu\nu} \,\nabla_\mu \overline{c} \, \nabla_\nu c\right].
  \end{split}
\end{align}
The corresponding perturbative expansion is similar to previous cases:
\begin{align}
  \begin{split}
    \mathcal{A}_\text{ghost} =& -\int d^4 x \, \sqrt{-g} \, g^{\mu\nu} \pd_\mu \overline{c} \,\pd_\nu c \\
    =&\sum\limits_{n=0}^\infty\int \cfrac{d^4 p_1}{(2\pi)^4}\cfrac{d^4 p_2}{(2\pi)^4} \prod\limits_{i=1}^n \cfrac{d^4 k_i}{(2\pi)^4} \, (2\pi)^4 \delta\left( p_1 + p_2 + \sum k_i \right) \, h_{\rho_1\sigma_1}(k_1) \cdots h_{\rho_n\sigma_n}(k_n)\\
    &\times\,\kappa^n\, \overline{c}(p_1) \left[ \left(\sqrt{-g} \, g^{\mu\nu} \right)^{\rho_1\sigma_1\cdots\rho_n\sigma_n} (p_1)_\mu (p_2)_\nu \right] c(p_2) .
  \end{split}
\end{align}
This expression results in the following ghost propagator
\begin{align}\label{Maxwell_ghost_propagator}
  \begin{gathered}
    \begin{fmffile}{Diag05}
      \begin{fmfgraph*}(30,30)
        \fmfleft{L}
        \fmfright{R}
        \fmf{dots}{L,R}
      \end{fmfgraph*}
    \end{fmffile}
  \end{gathered}
  \hspace{20pt} =  i\, \cfrac{-1}{p^2} \,,
\end{align}
and in the following interaction rule:
\begin{align}\label{rule_Gh_1}
  \nonumber \\
  \begin{gathered}
    \begin{fmffile}{FR_Gh_1}
      \begin{fmfgraph*}(40,40)
        \fmfleft{L1,L2}
        \fmfright{R1,R2}
        \fmf{dbl_wiggly}{L1,V}
        \fmf{dbl_wiggly}{L2,V}
        \fmf{dots}{R1,V}
        \fmf{dots}{R2,V}
        \fmffreeze
        \fmfdot{V}
        \fmf{dots}{L1,L2}
        \fmflabel{$p_1$}{R1}
        \fmflabel{$p_2$}{R2}
        \fmflabel{$\rho_1\sigma_1$}{L1}
        \fmflabel{$\rho_n\sigma_n$}{L2}
      \end{fmfgraph*}
    \end{fmffile}
  \end{gathered} = i \, \kappa^n \,\left(\sqrt{-g} \, g^{\mu\nu} \right)^{\rho_1\sigma_1\cdots\rho_n\sigma_n} \,I_{\mu\nu}{}^{\alpha\beta} (p_1)_\alpha (p_2)_\beta . \\ \nonumber
\end{align}

Let us highlight one more time, the ghosts discussed above are the standard Faddeev-Popov ghosts and should be treated accordingly. They do not appear in external states and mainly appear at the loop level. In the context of gravity, they are critical because of the following. In a given diagram a vertex describing the interaction between gravitons and vectors accounts for both physical and non-physical vector field polarizations. The coupling of Faddeev-Popov ghosts with gravity cancels out the energy contribution related to non-physical polarizations, making them necessary for the consistency of the theory.

\section{$SU(N)$ Yang-Mills}\label{section_SUNYM}

Let us turn to the discussion of the gravitational interaction of the $SU(N)$ Yang-Mills model. In the flat spacetime the $SU(N)$ Yang-Mills model is given by the following action:
\begin{align}\label{SUNYM_flat}
  \begin{split}
    \mathcal{A} =& \int d^4 x \left[ \overline{\psi} \left(i\, \widehat{\mathcal{D}} - m\right) \psi - \cfrac14\, F^a_{\mu\nu} \, F^{a\mu\nu}\right]\\
    =&\int d^4 x \left[ \overline{\psi} (i\,\widehat{\pd} -m )\psi - \cfrac14\,\left(f^a_{\mu\nu}\right)^2 + \gs \,\overline{\psi} \widehat{A}\psi -\gs\,f^{abc}\pd_\mu A^a_\nu \,A^{b\mu} A^{c\nu} - \cfrac14\,\gs^2 \,f^{amn}\,f^{aij} \,(A^m\!\!\cdot\!\! A^i) (A^n\!\!\cdot\!\! A^j) \right] .
  \end{split}
\end{align}
Here the fermion covariant derivative is defined as follows:
\begin{align}
  \mathcal{D}_\mu \psi = \pd_\mu \psi - i\,\gs\,A_\mu\,\psi .
\end{align}
Field tensor $F_{\mu\nu}$ reads
\begin{align}
  F_{\mu\nu} = \pd_\mu A_\nu - \pd_\nu A_\mu -i\,\gs [A_\mu , A_\nu] .
\end{align}
The gauge field $A_\mu$ takes value in $SU(N)$ algebra:
\begin{align}
  A_\mu = A^a_\mu \, T^a ,
\end{align}
where $T^a$ are generators. This gives the following expression of the field tensor components
\begin{align}
  F^a_{\mu\nu} = \pd_\mu A^a_\nu - \pd_\nu A^a_\mu + \gs\, f^{abc} \, A^b_\mu A^c_\nu \,.
\end{align}
Here $f^{abc}$ are the structure constants of the algebra:
\begin{align}
  [T^a, T^b] = i\, f^{abc} \, T^c\,.
\end{align}

Generalization of action \eqref{SUNYM_flat} for the case of curved spacetime is rather simple. One shall use the proper four-volume invariant and modify covariant derivatives to account for the curved geometry. This produces the following action:
\begin{align}
  \mathcal{A} = \int d^4 x \, \sqrt{-g} \left[ \,\overline{\psi} \left( i\, \mathfrak{e}_m{}^\mu\,\gamma^m\,  \mathcal{D}_\mu  - m  \right) \psi -\cfrac14\,F^a_{\mu\nu} \,F^{a\mu\nu} \right]. 
\end{align}
Here $\mathfrak{e}_m{}^\mu$ is a vierbein. The covariant derivative for fermions now reads
\begin{align}
  \mathcal{D}_\mu \psi = \nabla_\mu \psi - i\,\gs\,A_\mu \,\psi ,
\end{align}
with $\nabla_\mu$ begin the part accounting for the spacetime curvature via the spin connection. The field tensor $F_{\mu\nu}$ shall also account for the spacetime curvature, but because of its structure, it preserves the simple form:
\begin{align}
  \begin{split}
    F_{\mu\nu} &= \nabla_\mu A_\nu - \nabla_\nu A_\mu - i\,\gs\,[A_\mu,A_\nu] \\
    & = \pd_\mu A_\nu - \Gamma_{\mu\nu}^\sigma A_\sigma - \pd_\nu A_\mu + \Gamma_{\nu\mu}^\sigma A_\sigma - i\,\gs\,[A_\mu,A_\nu]\\
    &=\pd_\mu A_\nu - \pd_\nu A_\mu -i\,\gs [A_\mu , A_\nu]\,.
  \end{split}
\end{align}
Consequently, the $SU(N)$ Yang-Mills action in a curved spacetime reads:
\begin{align}\label{SUNYM_curve}
  \begin{split}
    \mathcal{A} = &\int d^4 x \sqrt{-g} \Bigg[ \overline{\psi} \left( i\, \mathfrak{e}_m{}^\mu \, \gamma^m \, \nabla_\mu- m \right) \psi - \cfrac14\, \left(f^a_{\mu\nu}\right)^2\\
      &+ \gs \,\overline{\psi} \left( \mathfrak{e}_m{}^\mu \gamma^m \right) \psi\,A_\mu -g^{\mu\nu} g^{\alpha\beta} \, \gs \,f^{abc} \pd_\mu A^a_\alpha A^b_\nu A^c_\beta -\cfrac14\,\gs^2\,f^{amn}\,f^{aij}\,g^{\mu\nu} g^{\alpha\beta} \,A^m_\mu\,A^i_\nu\,A^n_\alpha\,A^j_\beta \Bigg]\,.
  \end{split}
\end{align}

Perturbative quantization of kinetic parts of the action is discussed above, so we proceed with the derivation of Feynman's rules for the interaction sector. The perturbative expansion for the term describing the coupling of fermions to a gauge vector is given by the following expression:
\begin{align}\label{ffv_vertex}
  \begin{split}
    &\int d^4 x \sqrt{-g} \,\gs\, \overline{\psi} \left(\mathfrak{e}_m{}^\mu \gamma^m \right) \psi \, A_\mu \\
    &=\sum\limits_{n=0}^\infty \int \cfrac{d^4 p_1}{(2\pi)^4}\,\cfrac{d^4 p_2}{(2\pi)^4}\,\cfrac{d^4 k}{(2\pi)^4} \,\prod\limits_{i=0}^n \cfrac{d^4 k_i}{(2\pi)^4}\, (2\pi)^4\,\delta\left(p_1+p_2+k+\sum k_i \right) \,h_{\rho_1\sigma_1}(k_1)\cdots h_{\rho_n\sigma_n}(k_n) \\
    &\hspace{10pt}\times \, \kappa^n \, \overline{\psi}(p_2) \left[\gs \,\gamma^m\,T^a \, \left(\sqrt{-g}\,\mathfrak{e}_m{}^\mu\right)^{\rho_1\sigma_1\cdots\rho_n\sigma_n} \right] \psi(p_1) \, A^a_\mu(k)\,.
  \end{split}
\end{align}
This expression produces the following Feynman rule:
\begin{align}\label{rule_QQG}
  \nonumber \\
  \begin{gathered}
    \begin{fmffile}{FR_QQG}
      \begin{fmfgraph*}(40,40)
        \fmfleft{L1,L2,L3}
        \fmftop{T}
        \fmfbottom{B}
        \fmfright{R}
        \fmf{fermion}{B,V,T}
        \fmf{gluon}{V,R}
        \fmf{phantom}{V,L2}
        \fmfdot{V}
        \fmffreeze
        \fmf{dbl_wiggly}{L1,V}
        \fmf{dbl_wiggly}{L3,V}
        \fmf{dots}{L1,L3}
        \fmflabel{$\rho_1\sigma_1$}{L1}
        \fmflabel{$\rho_n\sigma_n$}{L3}
        \fmflabel{$\mu,a$}{R}
      \end{fmfgraph*}
    \end{fmffile}
  \end{gathered} \hspace{30pt} = i\, \kappa^n\, \gs\,\gamma^m\,T^a\,\left(\sqrt{-g} \,\mathfrak{e}_m{}^\mu \right)^{\rho_1\sigma_1\cdots\rho_n\sigma_n} . \\ \nonumber
\end{align}
The perturbative expansion for the term cubic in gauge vectors reads:
\begin{align}\label{vvv_vertex}
  \begin{split}
    &\int d^4 x \sqrt{-g} \, (-\gs)\,f^{abc}\, g^{\mu\nu} g^{\alpha\beta}\, \pd_\mu A^a_\alpha A^b_\nu A^c_\beta \\
    & = \sum\limits_{n=0}^\infty \int \cfrac{d^4 p_1}{(2\pi)^4} \, \cfrac{d^4 p_2}{(2\pi)^4} \, \cfrac{d^4 p_3}{(2\pi)^4} \,\prod\limits_{i=1}^n\cfrac{d^4 k_i}{(2\pi)^4}\, (2\pi)^4\,\delta\left(p_1+p_2+p_3+\sum k_i\right) \,h_{\rho_1\sigma_1}(k_1) \cdots h_{\rho_n\sigma_n}(k_n) \\
    &\hspace{10pt}\times\,\kappa^n\,\left[ (-i\, \gs) \,f^{abc} \, (p_1)_{\sigma}\, \left(\sqrt{-g} \,g^{\mu_1\mu_3} g^{\mu_2\sigma}\right)^{\rho_1\sigma_1\cdots\rho_n\sigma_n} \right] \, A^a_{\mu_1}(p_1)\,A^b_{\mu_2}(p_2)\,A^c_{\mu_3}(p_3) \,.
  \end{split}
\end{align}
This expression produces the following rule:
\begin{samepage}
  \begin{align*}
    \\
    \begin{gathered}
      \begin{fmffile}{FR_GGG}
        \begin{fmfgraph*}(60,40)
          \fmfleft{L1,L2,L3}
          \fmfright{R1,R2,R3}
          \fmf{dbl_wiggly,tension=2}{L1,V}
          \fmf{dbl_wiggly,tension=2}{L3,V}
          \fmf{gluon,tension=0.5}{V,R1}
          \fmf{gluon,tension=0.5}{V,R2}
          \fmf{gluon,tension=0.5}{V,R3}
          \fmffreeze
          \fmf{dots}{L1,L3}
          \fmfdot{V}
          \fmflabel{$\rho_1\sigma_1$}{L1}
          \fmflabel{$\rho_n\sigma_n$}{L3}
          \fmflabel{$\mu_1,a,p_1$}{R1}
          \fmflabel{$\mu_2,b,p_2$}{R2}
          \fmflabel{$\mu_3,c,p_3$}{R3}
        \end{fmfgraph*}
      \end{fmffile}
    \end{gathered}
  \end{align*}
  \begin{align}\label{rule_GGG}
    \begin{split}
      =& \kappa^n \,\gs\,f^{abc} \Big[ (p_1-p_2)_{\sigma}\left(\sqrt{-g} \,g^{\mu_1\mu_2} g^{\mu_3\sigma}\right)^{\rho_1\sigma_1\cdots\rho_n\sigma_n}  \\
        & +(p_3-p_1)_{\sigma} \left(\sqrt{-g} \,g^{\mu_1\mu_3} g^{\mu_2\sigma}\right)^{\rho_1\sigma_1\cdots\rho_n\sigma_n} +(p_2-p_3)_{\sigma} \left(\sqrt{-g} \,g^{\mu_2\mu_3} g^{\mu_1\sigma}\right)^{\rho_1\sigma_1\cdots\rho_n\sigma_n} \Big] .
    \end{split}
  \end{align}
\end{samepage}
Lastly, the term describing the four-vector coupling has the following perturbative expansion:
\begin{align}\label{vvvv_vertex}
  \begin{split}
    &\int d^4 x \sqrt{-g} \left(-\cfrac14\,\gs^2\right)\, f^{amn} f^{aij} \,g^{\mu\nu} g^{\alpha\beta} \, A^m_\mu \, A^i_\nu\, A^n_\alpha\, A^j_\beta\\
    &=\sum\limits_{n=0}^\infty\int \cfrac{d^4 p_1}{(2\pi)^4}\,\cfrac{d^4 p_2}{(2\pi)^4}\,\cfrac{d^4 p_3}{(2\pi)^4}\,\cfrac{d^4 p_4}{(2\pi)^4}\,\prod\limits_{i=0}^n \cfrac{d^4 k_i}{(2\pi)^4}\,(2\pi)^4 \,\delta\left(p_1+p_2+p_3+p_4+\sum k_i \right) h_{\rho_1\sigma_1}(k_1)\cdots h_{\rho_n\sigma_n}(k_n) \\
    &\hspace{10pt}\times\left( -\cfrac14 \right) \gs^2 \kappa^n f^{amn} f^{aij} \left(\sqrt{-g}\, g^{\mu_1\mu_3} g^{\mu_2\mu_4}\right)^{\rho_1\sigma_1\cdots\rho_n\sigma_n}  A^m_{\mu_1}(p_1)A^n_{\mu_2}(p_2) A^i_{\mu_3}(p_3)A^j_{\mu_4}(p_4) .
  \end{split}
\end{align}
This results in the following interaction rule:
\begin{align*}
  \\
  \begin{gathered}
    \begin{fmffile}{FR_GGGG}
      \begin{fmfgraph*}(50,50)
        \fmfleft{L1,L2}
        \fmfright{R1,R2,R3,R4}
        \fmf{gluon,tension=.5}{R1,V}
        \fmf{gluon,tension=.5}{R2,V}
        \fmf{gluon,tension=.5}{R3,V}
        \fmf{gluon,tension=.5}{R4,V}
        \fmf{dbl_wiggly,tension=2}{L1,V}
        \fmf{dbl_wiggly,tension=2}{L2,V}
        \fmfdot{V}
        \fmffreeze
        \fmf{dots}{L1,L2}
        \fmflabel{$\rho_1\sigma_1$}{L1}
        \fmflabel{$\rho_n\sigma_n$}{L2}
        \fmflabel{$\mu_1,a_1$}{R1}
        \fmflabel{$\mu_2,a_2$}{R2}
        \fmflabel{$\mu_3,a_3$}{R3}
        \fmflabel{$\mu_4,a_4$}{R4}
      \end{fmfgraph*}
    \end{fmffile}
  \end{gathered}
\end{align*}
\begin{align}\label{rule_GGGG}
  \begin{split}
    =  -i\,\gs^2 \kappa^n \Bigg[& f^{a_1 a_4 s} f^{a_2 a_3 s} \left( \left(\sqrt{-g}\, g^{\mu_1\mu_2}g^{\mu_3\mu_4}\right)^{\rho_1\cdots\sigma_n}-\left(\sqrt{-g}\, g^{\mu_1\mu_3}g^{\mu_2\mu_4}\right)\right)^{\rho_1\cdots\sigma_n} \\
      &+f^{a_1 a_3 s} f^{a_2 a_4 s} \left( \left(\sqrt{-g}\, g^{\mu_1\mu_2}g^{\mu_3\mu_4}\right)^{\rho_1\cdots\sigma_n}-\left(\sqrt{-g}\, g^{\mu_1\mu_4}g^{\mu_2\mu_3}\right)\right)^{\rho_1\cdots\sigma_n} \\
      & +f^{a_1 a_2 s} f^{a_3 a_4 s} \left( \left(\sqrt{-g}\, g^{\mu_1\mu_3}g^{\mu_2\mu_4}\right)^{\rho_1\cdots\sigma_n}-\left(\sqrt{-g}\, g^{\mu_1\mu_4}g^{\mu_2\mu_3}\right)\right)^{\rho_1\cdots\sigma_n} \Bigg].
  \end{split}
\end{align}

Finally, we shall turn to a discussion of the gauge fixing and the Faddeev-Popov ghosts. The Yang-Mills action \eqref{SUNYM_curve} respects the following gauge transformations:
\begin{align}
  \begin{split}
    \delta \psi =& i\,\theta^a \, T^a \psi ,\\
    \delta A_\mu =& i\,\theta^a \,[T^a, A_\mu] + \cfrac{1}{\gs} \, \pd_\mu \theta^a\,T^a ,\\
    \delta A^a_\mu =& \cfrac{1}{\gs} \left[\pd_\mu \theta^a - g \, f^{abc} \, \theta^b \, A^c_\mu \right] .
  \end{split}
\end{align}
Here $\theta^a$ are the gauge parameters. In the flat spacetime, one would use the standard Lorentz gauge fixing conditions
\begin{align}
  \pd^\mu A^a_\mu =0 .
\end{align}
For the curved spacetime case, the standard derivative shall be replaced with the covariant derivative, so the Lorentz gauge fixing conditions read:
\begin{align}
  g^{\mu\nu} \nabla_\mu A^a_\nu =0 .
\end{align}
We use this gauge fixing condition to introduce the Faddeev-Popov ghosts with the procedure discussed in the previous section. The introduction of this gauge fixing term will bring the kinetic part of the vector field to the same form obtained in the previous section. 

The ghost action is defined by the Faddeev-Popov determinant obtained from the gauge fixing condition:
\begin{align}
  \det\left[ \cfrac{\delta}{\delta \theta^b}\,\left\{ g^{\mu\nu} \nabla_\mu A^a_\nu \right\}  \right]  = \det\left[ \cfrac{1}{\gs}\,g^{\mu\nu} \, \nabla_\mu \left(\delta^{ab}\nabla_\nu  - \gs \,f^{abc}  \, A^c_\nu \right)\right].
\end{align}
It results in the following action:
\begin{align}
  \mathcal{A}_\text{FP} = \int d^4 x \left[  - g^{\mu\nu}\, \nabla_\mu\overline{c}^a \nabla_\nu c^a + \gs \,g^{\mu\nu} \nabla_\mu \overline{c}^a f^{abc} c^b A_\nu^c  \right].
\end{align}
The kinetic part of the action is similar to the case of a single massless vector field discussed in the previous section. The part of this action describing the interaction between ghosts, vectors, and gravitons admits the following perturbative expansion:
\begin{align}
  \begin{split}
    &\int d^4 x \sqrt{-g} \left[ \gs\,\pd_\mu \overline{c}^a \, f^{abc} \, c^b \, A^{c\,\mu} \right]\\
    =&\sum\limits_{n=0}^\infty \int \cfrac{d^4 p_1}{(2\pi)^4} \cfrac{d^4 p_2}{(2\pi)^4} \cfrac{d^4 k}{(2\pi)^4}\prod\limits_{i=0}^n \cfrac{d^4 k_i}{(2\pi)^4}\, (2\pi)^4 \delta\left(p_1+p_2+k+\sum k_i\right)\, h_{\rho_1\sigma_1}(k_1)\cdots h_{\rho_n\sigma_n}(k_n)\\
    &\times\, i\,\kappa^n\,\gs\,(p_1)_\nu \, f^{abc} \, \overline{c}^a(p_1) \,c^b(p_2) \, \left(\sqrt{-g}\,g^{\mu\nu}\right)^{\rho_1\sigma_1\cdots\rho_n\sigma_n}\,\,A^c_\mu(k).
  \end{split}
\end{align}
This expression produced the following rule:
\begin{align}\label{rule_GhGhG}
  \nonumber \\
  \begin{gathered}
    \begin{fmffile}{FR_GhGhG}
      \begin{fmfgraph*}(50,50)
        \fmfleft{L1,L2}
        \fmfright{R1,R2,R3}
        \fmf{dbl_wiggly,tension=2}{L1,V}
        \fmf{dbl_wiggly,tension=2}{L2,V}
        \fmf{gluon,tension=.5}{V,R2}
        \fmf{dots_arrow,tension=.5}{R1,V}
        \fmf{dots_arrow,tension=.5}{V,R3}
        \fmfdot{V}
        \fmffreeze
        \fmf{dots}{L1,L2}
        \fmflabel{$\rho_1\sigma_1$}{L1}
        \fmflabel{$\rho_n\sigma_n$}{L2}
        \fmflabel{$\mu,c$}{R2}
        \fmflabel{$b$}{R1}
        \fmflabel{$p_1,a$}{R3}
      \end{fmfgraph*}
    \end{fmffile}
  \end{gathered}
  \hspace{20pt}=& - \kappa^n\,g_s\,f^{abc} \, (p_1)_\nu \,\left(\sqrt{-g}\,g^{\mu\nu}\right)^{\rho_1\sigma_1\cdots\rho_n\sigma_n}. \\ \nonumber
\end{align}
Here all momenta are directed inwards and related by the conservation law.

Formulae \eqref{rule_F_1}, \eqref{rule_V_2}, \eqref{rule_Gh_1}, \eqref{rule_QQG}, \eqref{rule_GGG}, \eqref{rule_GGGG}, \eqref{rule_GhGhG} provide the complete set of Feynman rules required for treatment of the $SU(N)$ Yang-Mills model within perturbative quantum gravity.

\section{Faddeev-Popov ghosts for gravity}\label{section_gravity_ghosts}

General relativity is a gauge theory in the sense that it is invariant with respect to local transformations spawned by coordinate transformations. Therefore, a gauge fixing procedure similar to that for a gauge vector field shall be performed. However, the gravitational theory presents a more sophisticated system because of its geometrical nature. The perturbative approach to quantum gravity operates with small metric perturbations about the flat background. It may seem that this reduces gravity to a gauge theory of rank-$2$ symmetric tensor $h_{\mu\nu}$ propagating about the flat spacetime, but this is not the case.

A geometrical theory of gravity is different from a gauge theory of rank-$2$ symmetric tensor. The distinction lies in the choice of gauge fixing conditions. Let us begin with the case of a theory of a symmetric tensor $h_{\mu\nu}$ with the following gauge symmetry:
\begin{align}
  \delta h_{\mu\nu} = \pd_\mu \zeta_\nu + \pd_\nu \zeta_\mu .
\end{align}
The fundamental object of this theory is the $h_{\mu\nu}$ tensor, so a gauge fixing condition can be stated in terms of $h_{\mu\nu}$ alone. For instance,  one can use the following gauge fixing condition:
\begin{align}\label{naive_gauge_fixing}
  \pd_\mu h^{\mu\nu}  - \cfrac12\,\pd^\nu h = 0\,.
\end{align}
This (or any other) condition defines the structure of the Faddeev-Popov ghosts. Most importantly, since $h_{\mu\nu}$ is the fundamental object of the theory, the structure of divergencies can only be expressed in terms of $h_{\mu\nu}$ alone. Within the perturbative approach, all geometric quantities (Riemann tensor, Ricci tensor, scalar curvature, etc) are expressed in terms of small metric perturbations. The opposite is not true because there are operators given in terms of $h_{\mu\nu}$ alone that do not represent any geometric quantities. Consequently, within a gauge theory of a rank-$2$ symmetric tensor, one expects to find divergencies that cannot be described by geometric quantities and the theory can no longer be treated as a geometrical theory.

The situation is different for the consistent treatment of general relativity (or any other geometrical theory). Within the geometrical approach, gauge transformations are not introduced arbitrarily, but are related to coordinate frame transformations. This has two immediate implications. Firstly, within a geometrical theory gauge transformations are given by the so-called Lie derivatives:
\begin{align}
  \delta g_{\mu\nu} \overset{\text{def}}{=} \mathcal{L}_\zeta g_{\mu\nu} = \nabla_\mu \zeta_\nu + \nabla_\nu \zeta_\mu .
\end{align}
Here $\mathcal{L}_\zeta$ is the Lie derivative with respect to an arbitrary vector field $\zeta$ which plays the role of gauge parameters. Secondly, any suitable gauge fixing conditions must be expressed in terms of geometrical quantities. Because of this, gauge fixing conditions \eqref{naive_gauge_fixing} are inconsistent with the geometrical approach and they cannot be imposed. Instead, we use the following gauge fixing conditions:
\begin{align}\label{the_gravity_gauge_fixing}
  \mathcal{G}^\mu \overset{\text{note}}{=} g^{\alpha\beta} \Gamma^\mu_{\alpha\beta} =0 .
\end{align}
Together with the perturbative expansion \eqref{the_perturbative_expansion} gauge fixing conditions \eqref{the_gravity_gauge_fixing} spawns the following infinite series:
\begin{align}
  \mathcal{G}^\nu = \cfrac{\kappa}{2}\,g^{\mu\nu} g^{\alpha\beta} \left[ \pd_\alpha h_{\beta\mu} + \pd_\beta h_{\alpha\mu} - \pd_\mu h_{\alpha\beta}\right] =  \kappa \left[\pd_\mu h^{\mu\nu} - \cfrac12\,\pd^\nu h \right] + \mathcal{O}(\kappa^2) ,
\end{align}
The leading term of this series reproduces \eqref{naive_gauge_fixing}. Within the geometric theory, this series cannot be truncated, so the ghost sector is defined by the whole infinite expansion.

The need to use gauge fixing condition \eqref{the_gravity_gauge_fixing} except \eqref{naive_gauge_fixing} marks the difference between geometrical theories of gravity and a gauge theory of $h_{\mu\nu}$ tensor. The Faddeev-Popov prescription for the geometrical approach shall be constructed as follows. Firstly, we shall note that the gauge fixing condition $\mathcal{G}^\mu$ defined by \eqref{the_gravity_gauge_fixing} is a vector with mass dimension $+1$. Consequently, the general relativity action with the corresponding gauge fixing term shall be equipped with an additional dimensional parameter:
\begin{align}\label{Hilbert_gauge_fixed}
  \mathcal{A}_{\text{H}+\text{gf}} = \int d^4 x \sqrt{-g} \left[ -\cfrac{2}{\kappa^2}\,R + \cfrac{\epsilon}{2\,\kappa^2} \, g_{\mu\nu} \,\mathcal{G}^\mu \mathcal{G}^\nu \right].
\end{align}
Secondly, the corresponding Faddeev-Popov ghosts are also vectors. The structure of their action is defined by the variation of the gauge fixing term \eqref{the_gravity_gauge_fixing}:
\begin{align}
  \delta G^\mu = \mathcal{L}_\zeta \left[ g^{\alpha\beta} \, \Gamma^\mu_{\alpha\beta} \right] = \square \zeta^\mu - 2\, \Gamma^\mu_{\alpha\beta} \, \nabla^\alpha\zeta^\beta + R^\mu{}_\nu \zeta^\nu 
\end{align}
with $R_{\mu\nu}$ begin the Ricci tensor. Consequently, the ghost action reads:
\begin{align*}
  \mathcal{A}_\text{ghost} = \int d^4 x \sqrt{-g} \left[ -g^{\alpha\beta} g^{\mu\nu} \nabla_\alpha \overline{c}_\mu \nabla_\beta c_\nu - 2\,\Gamma^\mu_{\alpha\beta} \,\overline{c}_\mu \,\nabla^\alpha c^\beta + R_{\mu\nu} \, \overline{c}^\mu \,c^\nu \right].
\end{align*}
In the rest respect, the treatment of the Faddeev-Popov ghosts remains the same.

The structure of Feynman rules for gravity in this gauge is derived via the standard perturbative expansion. The structure of graviton interactions is given by action \eqref{Hilbert_gauge_fixed}:
\begin{align}
  \begin{split}
    \mathcal{A}_{\text{H}+\text{gf}} &= \int d^4 x \sqrt{-g} \left[ -\cfrac{2}{\kappa^2}\,R + \cfrac{\epsilon}{2\,\kappa^2}\, g_{\mu\nu}\, g^{\alpha\beta}\, g^{\rho\sigma} \,\Gamma^\mu_{\alpha\beta}\,\Gamma^\nu_{\rho\sigma}\right] = \\
    & = \int d^4 x \sqrt{-g} \, g^{\mu\nu} g^{\alpha\beta} g^{\rho\sigma} \left(-\cfrac{2}{\kappa^2}\right) \left[ \Gamma_{\alpha\mu\rho}\Gamma_{\sigma\nu\beta} - \Gamma_{\alpha\mu\nu} \Gamma_{\rho\beta\sigma} - \cfrac{\epsilon}{4} \,\Gamma_{\mu\alpha\beta} \Gamma_{\nu\rho\sigma} \right]\\
    & = -\cfrac12\,\int d^4 x \sqrt{-g}\,g^{\mu\nu} g^{\alpha\beta} g^{\rho\sigma} \Bigg[ \pd_\mu h_{\alpha\beta} \pd_\nu h_{\rho\sigma} - \pd_\mu h_{\alpha\rho} \pd_\nu h_{\beta\sigma} + 2\, \pd_\mu h_{\alpha\rho} \pd_\beta h_{\nu\sigma} -2\,\pd_\mu h_{\nu\alpha} \pd_\beta h_{\rho\sigma} \\
      & \hspace{170pt}  -\epsilon\, \left(\pd_\mu h_{\nu\rho} \pd_\alpha h_{\beta\sigma} - \pd_\mu h_{\alpha\beta} \pd_\rho h_{\sigma\nu} + \frac14\, \pd_\mu h_{\alpha\beta} \pd_\nu h_{\rho\sigma}  \right) \Bigg].
  \end{split}
\end{align}
It admits the following perturbative expansion:
\begin{align}
  \begin{split}
    \mathcal{A}_{\text{H}+\text{gf}} =& \sum\limits_{n=0}^\infty\int\cfrac{d^4 p_1}{(2\pi)^4} \cfrac{d^4 p_2}{(2\pi)^4} \prod\limits_{i=1}^n \cfrac{d^4 k_i}{(2\pi)^4}\,(2\pi)^4\,\delta\Big(p_1+p_2+\sum k_i \Big) \, h_{\rho_1\sigma_1}(k_1) \cdots h_{\rho_n\sigma_n}(k_n)\\
    &\times \left(2\,\kappa^n \right)\left( \sqrt{-g} \, g^{\mu\nu} g^{\alpha\beta} g^{\rho\sigma} \right)^{\rho_1\sigma_1\cdots\rho_n\sigma_n} (p_1)_{\lambda_1} (p_2)_{\lambda_2} \, h_{\mu_1\nu_1}(p_1) h_{\mu_2\nu_2}(p_2) \\
    & \times \Bigg[ \left(\Gamma_{\alpha\mu\rho}\right)^{\lambda_1\mu_1\nu_1} \left(\Gamma_{\sigma\nu\beta}\right)^{\lambda_2\mu_2\nu_2} - \left(\Gamma_{\alpha\mu\nu} \right)^{\lambda_1\mu_1\nu_1} \left( \Gamma_{\rho\beta\sigma}\right)^{\lambda_2\mu_2\nu_2} - \cfrac{\epsilon}{4} \left( \Gamma_{\mu\alpha\beta} \right)^{\lambda_1\mu_1\nu_1} \left(\Gamma_{\nu\rho\sigma} \right)^{\lambda_2\mu_2\nu_2}\Bigg].
  \end{split}
\end{align}
The complete expression for the graviton vertex is given by the following formula:
\begin{align*}
  \nonumber \\
  \begin{gathered}
    \begin{fmffile}{FR_h}
      \begin{fmfgraph*}(40,40)
        \fmfleft{L1,L2}
        \fmfright{R1,R2}
        \fmf{dbl_wiggly}{L1,V}
        \fmf{dbl_wiggly}{L2,V}
        \fmf{dbl_wiggly}{R1,V}
        \fmf{dbl_wiggly}{R2,V}
        \fmffreeze
        \fmfdot{V}
        \fmf{dots}{L1,L2}
        \fmflabel{$\mu_1\nu_1,p_1$}{R1}
        \fmflabel{$\mu_2\nu_2,p_2$}{R2}
        \fmflabel{$\mu_3\nu_3,p_3$}{L1}
        \fmflabel{$\mu_n\nu_n,p_n$}{L2}
      \end{fmfgraph*}
    \end{fmffile}
  \end{gathered}
\end{align*}
\begin{align}\label{Rule_Gravitons}
  \begin{split}
    = &\,i\,2\,\kappa^{n-2}\,\left(\sqrt{-g}\,g^{\mu\nu}g^{\alpha\beta}g^{\rho\sigma}\right)^{\mu_3\nu_3\cdots\mu_n\nu_n} \, (p_1)_{\lambda_1} (p_2)_{\lambda_2} \\
    & \times \Bigg[ \left(\Gamma_{\alpha\mu\rho}\right)^{\lambda_1\mu_1\nu_1} \left(\Gamma_{\sigma\nu\beta}\right)^{\lambda_2\mu_2\nu_2} - \left(\Gamma_{\alpha\mu\nu} \right)^{\lambda_1\mu_1\nu_1} \left( \Gamma_{\rho\beta\sigma}\right)^{\lambda_2\mu_2\nu_2} - \cfrac{\epsilon}{4} \left( \Gamma_{\mu\alpha\beta} \right)^{\lambda_1\mu_1\nu_1} \left(\Gamma_{\nu\rho\sigma} \right)^{\lambda_2\mu_2\nu_2}\Bigg] \\
    & + \text{permutations}.
  \end{split}
\end{align}
Here the summation is performed over all possible permutations of graviton parameters $\{\mu_i\,\nu_i\,p_i\}$.

The ghost action is treated similarly.
\begin{align}
  \begin{split}
    \mathcal{A}_\text{ghost} =& \int d^4 x \sqrt{-g} \left[ -g^{\alpha\beta} g^{\mu\nu} \nabla_\alpha \overline{c}_\mu \nabla_\beta c_\nu - 2\,\Gamma^\mu_{\alpha\beta} \,\overline{c}_\mu \,\nabla^\alpha c^\beta + R_{\mu\nu} \, \overline{c}^\mu \,c^\nu \right] \\
    =& \int d^4 x \sqrt{-g} \left[ -g^{\mu\nu} g^{\alpha\beta} \, \pd_\alpha\overline{c}_\mu \, \pd_\beta c_\nu \right]\\
    & + \int d^4 x \sqrt{-g}\, g^{\mu\alpha}g^{\nu\beta}g^{\rho\sigma} \left[ \Gamma_{\beta\rho\alpha} \pd_\sigma \overline{c}_\mu c_\nu - \Gamma_{\alpha\rho\beta}\,\overline{c}_\mu \pd_\sigma c_\nu + \pd_\rho\Gamma_{\sigma\alpha\beta} \,\overline{c}_\mu \,c_\nu - \pd_\alpha \Gamma_{\rho\beta\sigma} \overline{c}_\mu c_\nu \right]\\
    & + \int d^4 x \sqrt{-g}\,g^{\mu\alpha} g^{\nu\beta} g^{\rho\sigma} g^{\lambda\tau} \left[ \Gamma_{\rho\alpha\lambda} \Gamma_{\sigma\beta\tau} - \Gamma_{\rho\alpha\beta} \Gamma_{\sigma\lambda\tau} + \Gamma_{\alpha\rho\lambda} \Gamma_{\beta\sigma\tau} \right] \overline{c}_\mu c_\nu .
  \end{split}
\end{align}
It has the following perturbative expansion:
\begin{align}
  \begin{split}
    \mathcal{A}_\text{ghost} =& \sum\limits_{n=0}^\infty\int \cfrac{d^4 p_1}{(2\pi)^4} \cfrac{d^4 p_2}{(2\pi)^4} \prod\limits_{i=1}^n \cfrac{d^4 k_i}{(2\pi)^4} \, (2\pi)^4 \delta\big(p_1+p_2 + \sum k_i\big) h_{\rho_1\sigma_1}(k_1)\cdots h_{\rho_n\sigma_n}(k_n) \, \overline{c}_\mu (p_1) c_\nu(p_2)\\
    &\times \kappa^n \,\left(\sqrt{-g}\,g^{\mu\nu} g^{\alpha\beta} \right)^{\rho_1\sigma_1\cdots\rho_n\sigma_n} (p_1)_\alpha (p_2)_\beta\\
    +& \sum\limits_{n=1}^\infty\int \cfrac{d^4 p_1}{(2\pi)^4} \cfrac{d^4 p_2}{(2\pi)^4} \prod\limits_{i=1}^n \cfrac{d^4 k_i}{(2\pi)^4} \, (2\pi)^4 \delta\big(p_1+p_2 + \sum k_i\big) h_{\rho_1\sigma_1}(k_1)\cdots h_{\rho_n\sigma_n}(k_n) \, \overline{c}_\mu (p_1) c_\nu(p_2)\\
    &\times \kappa^n (-1)\left(\sqrt{-g}\,g^{\mu\alpha}g^{\nu\beta} g^{\rho\sigma}\right)^{\rho_2\sigma_2\cdots\rho_n\sigma_n} (k_1)_\lambda\left[ (p_1)_\sigma \left(\Gamma_{\beta\rho\alpha}\right)^{\lambda\rho_1\sigma_1}-(p_2)_\sigma \left(\Gamma_{\alpha\rho\beta}\right)^{\lambda\rho_1\sigma_1} \right.\\
      &\hspace{210pt}\left. + (k_1)_\rho \left(\Gamma_{\sigma\alpha\beta}\right)^{\lambda\rho_1\sigma_1} - (k_1)_\alpha \left(\Gamma_{\rho\beta\sigma}\right)^{\lambda\rho_1\sigma_1}\right]\\
    +& \sum\limits_{n=2}^\infty\int \cfrac{d^4 p_1}{(2\pi)^4} \cfrac{d^4 p_2}{(2\pi)^4} \prod\limits_{i=1}^n \cfrac{d^4 k_i}{(2\pi)^4} \, (2\pi)^4 \delta\big(p_1+p_2 + \sum k_i\big) h_{\rho_1\sigma_1}(k_1)\cdots h_{\rho_n\sigma_n}(k_n) \, \overline{c}_\mu (p_1) c_\nu(p_2)\\
    &\times \kappa^n (-1) \left(\sqrt{-g} \,g^{\mu\alpha}g^{\nu\beta} g^{\rho\sigma} g^{\lambda\tau}\right)^{\rho_3\sigma_3\cdots\rho_n\sigma_n}\,(k_1)_{\lambda_1} (k_2)_{\lambda_2} \left[ \left(\Gamma_{\rho\alpha\lambda}\right)^{\lambda_1\rho_1\sigma_1} \left(\Gamma_{\sigma\beta\tau}\right)^{\lambda_2\rho_2\sigma_2} \right.\\
      & \hspace{140pt}\left. - \left(\Gamma_{\rho\alpha\beta}\right)^{\lambda_1\rho_1\sigma_1} \left(\Gamma_{\sigma\lambda\tau}\right)^{\lambda_2\rho_2\sigma_2} + \left(\Gamma_{\alpha\rho\lambda}\right)^{\lambda_1\rho_1\sigma_1} \left(\Gamma_{\beta\sigma\tau}\right)^{\lambda_2\rho_2\sigma_2} \right].
  \end{split}
\end{align}
The complete expression describing graviton-ghost vertices reads:
\begin{samepage}
  \begin{align*}
    \\
    \begin{gathered}
      \begin{fmffile}{FR_Gh_2}
        \begin{fmfgraph*}(40,40)
          \fmfleft{L1,L2}
          \fmfright{R1,R2}
          \fmf{dbl_wiggly}{L1,V}
          \fmf{dbl_wiggly}{L2,V}
          \fmf{dots_arrow}{R1,V}
          \fmf{dots_arrow}{V,R2}
          \fmfdot{V}
          \fmffreeze
          \fmf{dots}{L1,L2}
          \fmflabel{$\rho_1\sigma_1,k_1$}{L1}
          \fmflabel{$\rho_n\sigma_n,k_n$}{L2}
          \fmflabel{$\nu,p_2$}{R1}
          \fmflabel{$\mu,p_1$}{R2}
        \end{fmfgraph*}
      \end{fmffile}
    \end{gathered}
  \end{align*}
  \begin{align}\label{Rules_Graviton-Ghosts}
    \begin{split}
      = i\,\kappa^n \Bigg[& \left(\sqrt{-g} \,g^{\mu\nu}g^{\alpha\beta}\right)^{\rho_1\sigma_1\cdots\rho_n\sigma_n} (p_1)_\alpha (p_2)_\beta \\
        &+\Bigg\{ - \left(\sqrt{-g}\,g^{\mu\alpha} g^{\nu\beta}g^{\rho\sigma}\right)^{\rho_2\sigma_2\cdots\rho_n\sigma_n}(k_1)_\lambda \Bigg[ (p_1)_\sigma (\Gamma_{\beta\rho\alpha})^{\lambda\rho_1\sigma_1} - (p_2)_\sigma (\Gamma_{\alpha\rho\beta})^{\lambda\rho_1\sigma_1} \\
          & \hspace{150pt}+ (k_1)_\rho (\Gamma_{\sigma\alpha\beta})^{\lambda\rho_1\sigma_1}-(k_1)_\alpha (\Gamma_{\rho\beta\sigma})^{\lambda\rho_1\sigma_1} \Bigg] + \text{permutations} \Bigg\}\\
        & + \Bigg\{ -\left(\sqrt{-g}\,g^{\mu\alpha} g^{\nu\beta}g^{\rho\sigma} g^{\lambda\tau} \right)^{\rho_3\sigma_3\cdots\rho_n\sigma_n} \,(k_1)_{\lambda_1} (k_2)_{\lambda_2} \left[ \left(\Gamma_{\rho\alpha\lambda}\right)^{\lambda_1\rho_1\sigma_1} \left(\Gamma_{\sigma\beta\tau}\right)^{\lambda_2\rho_2\sigma_2} \right.\\
          & \hspace{70pt}\left. - \left(\Gamma_{\rho\alpha\beta}\right)^{\lambda_1\rho_1\sigma_1} \left(\Gamma_{\sigma\lambda\tau}\right)^{\lambda_2\rho_2\sigma_2} + \left(\Gamma_{\alpha\rho\lambda}\right)^{\lambda_1\rho_1\sigma_1} \left(\Gamma_{\beta\sigma\tau}\right)^{\lambda_2\rho_2\sigma_2} \right] + \text{permutations} \Bigg\} \Bigg].
    \end{split}
  \end{align}
\end{samepage}

Propagators for ghosts and gravitons are derived by the standard procedure. The ghost propagator is given by the following expression.
\begin{align}
  \begin{gathered}
    \begin{fmffile}{FR_Ghost_Propagator}
      \begin{fmfgraph*}(30,30)
        \fmfleft{L}
        \fmfright{R}
        \fmf{dots}{L,R}
        \fmflabel{$\mu$}{L}
        \fmflabel{$\nu$}{R}
      \end{fmfgraph*}
    \end{fmffile}
  \end{gathered}
  \hspace{20pt} = i \, \cfrac{\eta_{\mu\nu}}{k^2}\,.
\end{align}
The graviton propagator contains the gauge fixing parameter $\epsilon$. The propagator corresponds to the part of the microscopic action quadratic in perturbations:
\begin{align}
  \int d^4 x \sqrt{-g} \left[ -\cfrac{2}{\kappa^2}\, R + \cfrac{\epsilon}{2\,\kappa^2} ~\mathcal{G}_\mu \mathcal{G}^\mu \right] = \int d^4 x \left[  -\cfrac12\,h^{\mu\nu} \mathcal{D}_{\mu\nu\alpha\beta}(\epsilon)\, \square h^{\alpha\beta}  \right] +\korder{1}.
\end{align}
In the momentum representation the operator $\mathcal{D}$ is given in terms of the Nieuwenhuizen operators \cite{VanNieuwenhuizen:1981ae,Accioly:2000nm}
\begin{align}
  \mathcal{D}_{\mu\nu\alpha\beta} (\epsilon) = \cfrac{3 \epsilon -8}{4}\, P^0_{\mu\nu\alpha\beta} + \cfrac{\epsilon}{2}\,P^1_{\mu\nu\alpha\beta} + P^2_{\mu\nu\alpha\beta} - \cfrac{\epsilon}{4} ~ \overline{P}^0_{\mu\nu\alpha\beta} - \cfrac{\epsilon}{4} ~ \overline{\overline{P}}^0_{\mu\nu\alpha\beta} \,.
\end{align}
Only $P^0$ and $P^2$ operators are gauge invariant, so the operator is invertible if $\epsilon \not =0$. The inverse operator reads:
\begin{align}
  \mathcal{D}^{-1}_{\mu\nu\alpha\beta}(\epsilon) = -\cfrac12\,P^0_{\mu\nu\alpha\beta}+\cfrac{2}{\epsilon}\, P^1_{\mu\nu\alpha\beta} + P^2_{\mu\nu\alpha\beta} - \cfrac{3\,\epsilon -8}{2\,\epsilon}~\overline{P}^0_{\mu\nu\alpha\beta} - \cfrac{1}{2} ~\overline{\overline{P}}^0_{\mu\nu\alpha\beta} .
\end{align}
Therefore, in an arbitrary gauge the graviton propagator is given by the following expression:
\begin{align}
  \begin{gathered}
    \begin{fmffile}{FR_Graviton_Propagator}
      \begin{fmfgraph*}(40,40)
        \fmfleft{L}
        \fmfright{R}
        \fmf{dbl_wiggly}{L,R}
        \fmflabel{$\mu\nu$}{L}
        \fmflabel{$\alpha\beta$}{R}
      \end{fmfgraph*}
    \end{fmffile}
  \end{gathered}
  \hspace{20pt} = i ~ \cfrac{ \mathcal{D}^{-1}_{\mu\nu\alpha\beta}(\epsilon) }{k^2} \, .
\end{align}
We will consider the general case within this paper. However, on the practical ground, the simplest choice of the gauge fixing parameter is $\epsilon =2$. With this value of the gauge fixing parameter the operator $\mathcal{D}^{-1}$ takes an extremely simple form:
\begin{align}
  \mathcal{D}^{-1}_{\mu\nu\alpha\beta}(2) = \cfrac12\left[ \eta_{\mu\alpha} \eta_{\nu\beta} + \eta_{\mu\beta} \eta_{\nu\alpha} - \eta_{\mu\nu} \eta_{\alpha\beta} \right].
\end{align}

\section{FeynGrav v2}\label{section_FG2}

FeynGrav, a package for computing Feynman rules for gravity, has been updated with new features and improvements. The latest version includes support for the interaction rules presented in the sections above, as well as additional capabilities to enhance its functionality. The code is publicly available \cite{FeynGrav}.

Firstly, the package structure has been changed. The main file ``FeynGrav.wl'' contains the code providing tools to operate with Feynman rules for gravity. The package is based on FeynCalc and requires it to run \cite{Mertig:1990an,Shtabovenko:2016sxi,Shtabovenko:2020gxv}. The package operates with pre-generated libraries which contain data on gravitational interaction. The folder ``Rules'' contains realizations of both interaction rules and supplementary functions. The folder ``Libs'' contains files with evaluated expressions for the interaction rules. The folder also contains a script ``FeynGravLibrariesGenerator.wl'' which generates those libraries. FeynGrav is distributed with libraries for gravitational interaction up to $\korder{3}$ order. The previous version of FeynGrav generated expressions for interaction vertices on a user's call which negatively affected the performance.

Secondly, the package is distributed with an example file ``FeynGrav\_Examples.nb''. The file contains the following examples:
\begin{list}{$\bullet$}{}
\item
  Realization of the Nieuwenhuizen operators \cite{VanNieuwenhuizen:1981ae,Accioly:2000nm};
\item
  Calculation of matrix element for on-shell tree-level $2\to 2$ graviton scattering that agrees with \cite{Sannan:1986tz}.
\item
  Calculation of various contributions to the graviton self-energy at the one-loop level.
\item
  One-loop matter polarization operators induced by gravity.
\item
  One-loop vertex function for a graviton-scalar vertex.
\end{list}

Thirdly, the package contains a few supplementary functions that are often used in quantum gravity calculations. These are propagator for a scalar field \eqref{scalar_propagator}, propagator for the Proca field (massive vector field) \eqref{Proca_propagator}, and the Nieuwenhuizen operators \cite{VanNieuwenhuizen:1981ae,Accioly:2000nm}. The Nieuwenhuizen operators are a generalization of the standard gauge projectors and they are discussed in many other publications, so we will not discuss them further. It must be noted that the original Nieuwenhuizen operators are defined in $d=4$ where they have a few special features. Within FeynGrav these operators are given in arbitrary $d$. This is done for the sake of consistency as most parts of tools provided by FeynCalc are designed to operate with arbitrary $d$.

The new version of FeynGrav also includes interaction rules for matter with $s=0$,$1$, and $1/2$ of arbitrary mass, as well as interaction rules for $SU(N)$ Yang-Mills that are consistent with the realization used within FeynCalc. The complete list of commands for interaction rules is given in Appendix \ref{Command_list}.

Lastly, the gravitational sector of the new FeynGrav version supports an arbitrary gauge fixing parameter present in \eqref{Hilbert_gauge_fixed}. The package is initiated with the corresponding parameter being unspecified and entering all expressions as a constant. At any point, the user is free to fix this parameter and proceed with the calculations. As was noted before, from the practical point of view $\epsilon=2$ gauge is the simplest because the graviton propagator takes a much simpler form.

All other features of FeynGrav remained unchanged from the previous version. They are described in detail in the previous paper \cite{Latosh:2022ydd}, so we will not discuss them further. However, we present some calculations that provide a suitable illustration of FeynGrav's applicability.

\subsection{Example of polarization operators}\label{section_physics}

To demonstrate the applicability of FeynGrav, we can use it to perform some typical quantum field theory calculations. Let's start with the calculation of various contributions to the graviton self-energy. In the following calculations, we express all loop integrals in terms of the Passarino-Veltman integrals \cite{Passarino:1978jh}. Since the calculations are performed within FeynCalc, we omit all $A_0(0)$ integrals but preserve $A_0(m^2)$ integrals.

Graviton polarization operator induced by a single scalar field:
\begin{align}
  \begin{split}
    i\, \Pi_{\mu\nu\alpha\beta}^{s=0,m_\text{s}} (p) =& \hspace{20pt}
    \begin{gathered}
      \begin{fmffile}{Loop_Scalar_1}
        \begin{fmfgraph*}(30,40)
          \fmfleft{L}
          \fmfright{R}
          \fmftop{T}
          \fmflabel{$\mu\nu$}{L}
          \fmflabel{$\alpha\beta$}{R}
          \fmf{dbl_wiggly}{L,V,R}
          \fmf{phantom}{V,T}
          \fmffreeze
          \fmf{dashes,right=1}{V,T,V}
          \fmfdot{V}
        \end{fmfgraph*}
      \end{fmffile}
    \end{gathered}
    \hspace{20pt}+\hspace{20pt}
    \begin{gathered}
      \begin{fmffile}{Loop_Scalar_2}
        \begin{fmfgraph*}(30,30)
          \fmfleft{L}
          \fmfright{R}
          \fmflabel{$\mu\nu$}{L}
          \fmflabel{$\alpha\beta$}{R}
          \fmf{dbl_wiggly,tension=2}{L,VL}
          \fmf{dbl_wiggly,tension=2}{VR,R}
          \fmf{dashes,right=1,tension=.5}{VL,VR,VL}
          \fmfdot{VL,VR}
        \end{fmfgraph*}
      \end{fmffile}
    \end{gathered}\\
    =& \kappa^2 ~ i \, \pi^2 B_0(p^2,m_\text{s}^2,m_\text{s}^2) \left[\cfrac{1}{12} \left(p^2+2\,m_\text{s}^2\right)^2 P^0_{\mu\nu\alpha\beta} +\cfrac{1}{120} \left(p^2-4\,m_\text{s}^2\right)^2 P^2_{\mu\nu\alpha\beta} \right] \\
    & -\kappa^2 ~ i\,\pi^2  A_0(m_\text{s}^2) \left[\cfrac{1}{6} \left( p^2 + 2\,m_\text{s}^2 \right) P^0_{\mu\nu\alpha\beta}  + \cfrac{1}{60} \left( p^2 + 8\,m_\text{s}^2 \right) P^2_{\mu\nu\alpha\beta} \right].
  \end{split}
\end{align}
Graviton polarization operator induced by a single Dirac field:
\begin{align}
  \begin{split}
    & i\, \Pi^{s=1/2,m_\text{f}}_{\mu\nu\alpha\beta}(p)= \hspace{20pt}
    \begin{gathered}
      \begin{fmffile}{Loop_Fermion_1}
        \begin{fmfgraph*}(30,30)
          \fmfleft{L}
          \fmfright{R}
          \fmftop{T}
          \fmflabel{$\mu\nu$}{L}
          \fmflabel{$\alpha\beta$}{R}
          \fmf{dbl_wiggly}{L,V,R}
          \fmf{phantom}{V,T}
          \fmffreeze
          \fmf{fermion,right=1}{V,T,V}
          \fmfdot{V}
        \end{fmfgraph*}
      \end{fmffile}
    \end{gathered}
    \hspace{20pt}+\hspace{20pt}
    \begin{gathered}
      \begin{fmffile}{Loop_Fermion_2}
        \begin{fmfgraph*}(30,30)
          \fmfleft{L}
          \fmfright{R}
          \fmflabel{$\mu\nu$}{L}
          \fmflabel{$\alpha\beta$}{R}
          \fmf{dbl_wiggly,tension=2}{L,VL}
          \fmf{dbl_wiggly,tension=2}{VR,R}
          \fmf{fermion,right=1,tension=.5}{VL,VR,VL}
          \fmfdot{VL,VR}
        \end{fmfgraph*}
      \end{fmffile}
    \end{gathered} \\
    &= \kappa^2\, i \,\pi^2\, B_0(p^2,m_\text{f}^2,m_\text{f}^2)\,(p^2- 4\,m_\text{f}^2)\, \left[ \cfrac16\,m_\text{f}^2\, P^0_{\mu\nu\alpha\beta}+\cfrac{1}{120}\, (3\,p^2+8\,m_\text{f}^2)\, P^2_{\mu\nu\alpha\beta}  \right]\\
    &\hspace{20pt} +\kappa^2\,i\,\pi^2\, A_0(m_\text{f}^2)  \left[ \cfrac{19}{24}\,m_\text{f}^2\, P^0_{\mu\nu\alpha\beta} +\cfrac{1}{8}\,m_\text{f}^2 P^1_{\mu\nu\alpha\beta} -\cfrac{1}{120}\,(6\,p^2-47\,m_\text{f}^2) \,P^2_{\mu\nu\alpha\beta} + \cfrac{m_\text{f}^2}{8}\, \overline{P}^0_{\mu\nu\alpha\beta} \right].
  \end{split}
\end{align}
Graviton polarization operator induced by a single Proca field:
\begin{align}
  \begin{split}
    &i\, \Pi^{s=1,m_\text{v}\not=0}_{\mu\nu\alpha\beta} = \hspace{20pt}
    \begin{gathered}
      \begin{fmffile}{Loop_Proca_1}
        \begin{fmfgraph*}(30,30)
          \fmfleft{L}
          \fmfright{R}
          \fmftop{T}
          \fmflabel{$\mu\nu$}{L}
          \fmflabel{$\alpha\beta$}{R}
          \fmf{dbl_wiggly}{L,V,R}
          \fmf{phantom}{V,T}
          \fmffreeze
          \fmf{photon,right=1}{V,T,V}
          \fmfdot{V}
        \end{fmfgraph*}
      \end{fmffile}
    \end{gathered}
    \hspace{20pt}+\hspace{20pt}
    \begin{gathered}
      \begin{fmffile}{Loop_Proca_2}
        \begin{fmfgraph*}(30,30)
          \fmfleft{L}
          \fmfright{R}
          \fmflabel{$\mu\nu$}{L}
          \fmflabel{$\alpha\beta$}{R}
          \fmf{dbl_wiggly,tension=2}{L,VL}
          \fmf{dbl_wiggly,tension=2}{VR,R}
          \fmf{photon,right=1,tension=.5}{VL,VR,VL}
          \fmfdot{VL,VR}
        \end{fmfgraph*}
      \end{fmffile}
    \end{gathered}
    \\
    &= \kappa^2 ~i\,\pi^2\,B_0(p^2,m_\text{v}^2,m_\text{v}^2) \left[ \cfrac{1}{12}\left( p^4 -4 \,m_\text{v}^2\,p^2 +12\,m_\text{v}^4 \right)\,P^0_{\mu\nu\alpha\beta} +\cfrac{1}{120}\left( 13\,p^4 +56\,m_\text{v}^2\,p^2+48\,m_\text{v}^4 \right)\, P^2_{\mu\nu\alpha\beta} \right] \\
    & \hspace{10pt} +\kappa^2~i\,\pi^2\, A_0(m_\text{v}^2) \left[ -\cfrac16\,(p^2 + 3 \,m_\text{v}^2) P^0_{\mu\nu\alpha\beta} - \cfrac{m_\text{v}^2}{4}\,P^1_{\mu\nu\alpha\beta} -\cfrac{13}{60}\,(p^2+3\,m_\text{v}^2)\,P^2_{\mu\nu\alpha\beta}+\cfrac{m_\text{v}^2}{4} ~ \overline{\overline{P}}^0_{\mu\nu\alpha\beta} \right].
  \end{split}
\end{align}
Graviton polarization operator induced by a single massless vector field ($\epsilon_\text{V}=-1$):
\begin{align}
  \begin{split}
    i\, \Pi^{s=1,m=0}_{\mu\nu\alpha\beta} =& \hspace{20pt}
    \begin{gathered}
      \begin{fmffile}{Loop_Maxwell_1}
        \begin{fmfgraph*}(30,30)
          \fmfleft{L}
          \fmfright{R}
          \fmftop{T}
          \fmflabel{$\mu\nu$}{L}
          \fmflabel{$\alpha\beta$}{R}
          \fmf{dbl_wiggly}{L,V,R}
          \fmf{phantom}{V,T}
          \fmffreeze
          \fmf{photon,right=1}{V,T,V}
          \fmfdot{V}
        \end{fmfgraph*}
      \end{fmffile}
    \end{gathered}
    \hspace{20pt}+\hspace{20pt}
    \begin{gathered}
      \begin{fmffile}{Loop_Maxwell_2}
        \begin{fmfgraph*}(30,30)
          \fmfleft{L}
          \fmfright{R}
          \fmftop{T}
          \fmflabel{$\mu\nu$}{L}
          \fmflabel{$\alpha\beta$}{R}
          \fmf{dbl_wiggly}{L,V,R}
          \fmf{phantom}{V,T}
          \fmffreeze
          \fmf{dots,right=1}{V,T,V}
          \fmfdot{V}
        \end{fmfgraph*}
      \end{fmffile}
    \end{gathered}
    \hspace{20pt}+\hspace{20pt}
    \begin{gathered}
      \begin{fmffile}{Loop_Maxwell_3}
        \begin{fmfgraph*}(30,30)
          \fmfleft{L}
          \fmfright{R}
          \fmflabel{$\mu\nu$}{L}
          \fmflabel{$\alpha\beta$}{R}
          \fmf{dbl_wiggly,tension=2}{L,VL}
          \fmf{dbl_wiggly,tension=2}{VR,R}
          \fmf{photon,right=1,tension=.5}{VL,VR,VL}
          \fmfdot{VL,VR}
        \end{fmfgraph*}
      \end{fmffile}
    \end{gathered}
    \hspace{20pt}+\hspace{20pt}
    \begin{gathered}
      \begin{fmffile}{Loop_Maxwell_4}
        \begin{fmfgraph*}(30,30)
          \fmfleft{L}
          \fmfright{R}
          \fmflabel{$\mu\nu$}{L}
          \fmflabel{$\alpha\beta$}{R}
          \fmf{dbl_wiggly,tension=2}{L,VL}
          \fmf{dbl_wiggly,tension=2}{VR,R}
          \fmf{dots,right=1,tension=.5}{VL,VR,VL}
          \fmfdot{VL,VR}
        \end{fmfgraph*}
      \end{fmffile}
    \end{gathered}
    \\
    = &  \kappa^2 ~i\,\pi^2\,B_0(p^2,0 ,0) \, p^2\,\left[ \frac{1}{4}\,P^0_{\mu\nu\alpha\beta} +\cfrac{1}{8}\, P^2_{\mu\nu\alpha\beta}  \right].
  \end{split}
\end{align}
Graviton polarization operator induced by $SU(N)$ Yang-Mills matter ($\epsilon_\text{SU(N)YM}=-1$):
\begin{align}
  \begin{split}
    i\, \Pi^{SU(N)\text{ Yang-Mills}}_{\mu\nu\alpha\beta} =& \hspace{20pt}
    \begin{gathered}
      \begin{fmffile}{Loop_SUNYM_1}
        \begin{fmfgraph*}(30,30)
          \fmfleft{L}
          \fmfright{R}
          \fmftop{T}
          \fmflabel{$\mu\nu$}{L}
          \fmflabel{$\alpha\beta$}{R}
          \fmf{dbl_wiggly}{L,V,R}
          \fmf{phantom}{V,T}
          \fmffreeze
          \fmf{gluon,right=.5}{V,T,V}
          \fmfdot{V}
        \end{fmfgraph*}
      \end{fmffile}
    \end{gathered}
    \hspace{20pt}+\hspace{20pt}
    \begin{gathered}
      \begin{fmffile}{Loop_SUNYM_2}
        \begin{fmfgraph*}(30,30)
          \fmfleft{L}
          \fmfright{R}
          \fmftop{T}
          \fmflabel{$\mu\nu$}{L}
          \fmflabel{$\alpha\beta$}{R}
          \fmf{dbl_wiggly}{L,V,R}
          \fmf{phantom}{V,T}
          \fmffreeze
          \fmf{dots,right=1}{V,T,V}
          \fmfdot{V}
        \end{fmfgraph*}
      \end{fmffile}
    \end{gathered}
    \hspace{20pt}+\hspace{20pt}
    \begin{gathered}
      \begin{fmffile}{Loop_SUNYM_3}
        \begin{fmfgraph*}(30,30)
          \fmfleft{L}
          \fmfright{R}
          \fmftop{T}
          \fmflabel{$\mu\nu$}{L}
          \fmflabel{$\alpha\beta$}{R}
          \fmf{dbl_wiggly}{L,V,R}
          \fmf{phantom}{V,T}
          \fmffreeze
          \fmf{fermion,right=1}{V,T,V}
          \fmfdot{V}
        \end{fmfgraph*}
      \end{fmffile}
    \end{gathered}
    \\
    & + \hspace{20pt}
    \begin{gathered}
      \begin{fmffile}{Loop_SUNYM_4}
        \begin{fmfgraph*}(30,30)
          \fmfleft{L}
          \fmfright{R}
          \fmflabel{$\mu\nu$}{L}
          \fmflabel{$\alpha\beta$}{R}
          \fmf{dbl_wiggly,tension=2}{L,VL}
          \fmf{dbl_wiggly,tension=2}{VR,R}
          \fmf{gluon,right=.5,tension=.5}{VL,VR,VL}
          \fmfdot{VL,VR}
        \end{fmfgraph*}
      \end{fmffile}
    \end{gathered}
    \hspace{20pt} + \hspace{20pt}
    \begin{gathered}
      \begin{fmffile}{Loop_SUNYM_5}
        \begin{fmfgraph*}(30,30)
          \fmfleft{L}
          \fmfright{R}
          \fmflabel{$\mu\nu$}{L}
          \fmflabel{$\alpha\beta$}{R}
          \fmf{dbl_wiggly,tension=2}{L,VL}
          \fmf{dbl_wiggly,tension=2}{VR,R}
          \fmf{dots,right=1,tension=.5}{VL,VR,VL}
          \fmfdot{VL,VR}
        \end{fmfgraph*}
      \end{fmffile}
    \end{gathered}
    \hspace{20pt} + \hspace{20pt}
    \begin{gathered}
      \begin{fmffile}{Loop_SUNYM_6}
        \begin{fmfgraph*}(30,30)
          \fmfleft{L}
          \fmfright{R}
          \fmflabel{$\mu\nu$}{L}
          \fmflabel{$\alpha\beta$}{R}
          \fmf{dbl_wiggly,tension=2}{L,VL}
          \fmf{dbl_wiggly,tension=2}{VR,R}
          \fmf{fermion,right=1,tension=.5}{VL,VR,VL}
          \fmfdot{VL,VR}
        \end{fmfgraph*}
      \end{fmffile}
    \end{gathered}
    \\
    =&   \kappa^2 ~i\,\pi^2\,B_0(p^2,0 ,0)\,p^4 \,\left[\cfrac{N^2-1}{4} \,P^0_{\mu\nu\alpha\beta} +\left( \cfrac{N^2-1}{8} - \cfrac{N}{40} \right) P^2_{\mu\nu\alpha\beta}  \right].
  \end{split}
\end{align}
Here $N$ is the number of colors.

In the same way, the graviton self-energy is calculated:
\begin{align}
  \begin{split}
    i\,\Pi_{\mu\nu\alpha\beta}^{s=2,m=0}&= \hspace{20pt}
    \begin{gathered}
      \begin{fmffile}{Loop_Gravity_0}
        \begin{fmfgraph*}(40,40)
          \fmfleft{L}
          \fmfright{R}
          \fmftop{T}
          \fmf{dbl_wiggly}{L,V,R}
          \fmf{phantom}{V,T}
          \fmffreeze
          \fmf{dbl_wiggly,right=1}{V,T,V}
          \fmfdot{V}
          \fmflabel{$\mu\nu$}{L}
          \fmflabel{$\alpha\beta$}{R}
        \end{fmfgraph*}
      \end{fmffile}
    \end{gathered}
    \hspace{20pt} + \hspace{20pt}
    \begin{gathered}
      \begin{fmffile}{Loop_Gravity_1}
        \begin{fmfgraph*}(40,40)
          \fmfleft{L}
          \fmfright{R}
          \fmf{dbl_wiggly}{L,VL}
          \fmf{dbl_wiggly}{VR,R}
          \fmf{dbl_wiggly,right=1,tension=.3}{VL,VR,VL}
          \fmflabel{$\mu\nu$}{L}
          \fmflabel{$\alpha\beta$}{R}
          \fmfdot{VL,VR}
        \end{fmfgraph*}
      \end{fmffile}
    \end{gathered}
    \hspace{20pt} + \hspace{20pt}
    \begin{gathered}
      \begin{fmffile}{Loop_Gravity_2}
        \begin{fmfgraph*}(40,40)
          \fmfleft{L}
          \fmfright{R}
          \fmf{dbl_wiggly}{L,VL}
          \fmf{dbl_wiggly}{VR,R}
          \fmf{dots,right=1,tension=.3}{VL,VR,VL}
          \fmflabel{$\mu\nu$}{L}
          \fmflabel{$\alpha\beta$}{R}
          \fmfdot{VL,VR}
        \end{fmfgraph*}
      \end{fmffile}
    \end{gathered}
    \\
    &=\kappa^2\,i\,\pi^2\,B_0(p^2,0,0)\,p^4\,\Bigg[\cfrac{13}{12}\,P^1_{\mu\nu\alpha\beta} - \cfrac{11}{20}\,P^2_{\mu\nu\alpha\beta} +\cfrac{7}{8}\,P^0_{\mu\nu\alpha\beta} +\cfrac{17}{8}\,\overline{P}^0_{\mu\nu\alpha\beta} -\cfrac{11}{24}\,\overline{\overline{P}}^0_{\mu\nu\alpha\beta} \Bigg].
  \end{split}
\end{align}

Gravitational contribution to matter propagators is given by the following expression in $d=4$ with the gauge fixing parameter $\epsilon=2$. The polarization operator for a scalar field reads:
\begin{align}
  \begin{split}
    i\,\Pi^{s=0,m_\text{s}} = &\hspace{10pt}
    \begin{gathered}
      \begin{fmffile}{Loop_Scalar_3}
        \begin{fmfgraph*}(30,30)
          \fmfleft{L}
          \fmfright{R}
          \fmftop{T}
          \fmf{dashes}{L,V,R}
          \fmf{phantom}{V,T}
          \fmffreeze
          \fmf{dbl_wiggly,right=1}{V,T,V}
          \fmfdot{V}
        \end{fmfgraph*}
      \end{fmffile}
    \end{gathered}
    \hspace{5pt} + \hspace{5pt}
    \begin{gathered}
      \begin{fmffile}{Loop_Scalar_4}
        \begin{fmfgraph*}(30,30)
          \fmfleft{L}
          \fmfright{R}
          \fmf{dashes,tension=2}{L,VL}
          \fmf{dashes,tension=2}{VR,R}
          \fmf{dashes,tension=.1}{VL,VR}
          \fmf{dbl_wiggly,left=1}{VL,VR}
          \fmfdot{VL,VR}
        \end{fmfgraph*}
      \end{fmffile}
    \end{gathered}
    \\
    =& -\cfrac{1}{2}\,m_\text{s}^2\,\kappa^2\,i\,\pi^2\,A_0(m_\text{s}^2) + m^2\,\left(p^2 - \cfrac{m_\text{s}^2}{2}\right)\,\kappa^2\,i\,\pi^2\, B_0(p^2,0,m_\text{s}^2).
  \end{split}
\end{align}
Polarization operator for a Dirac field:
\begin{align}
  \begin{split}
    i\,\Pi^{s=1/2,m_\text{f}} =& \hspace{10pt}
    \begin{gathered}
      \begin{fmffile}{Loop_Fermion_3}
        \begin{fmfgraph*}(40,40)
          \fmfleft{L}
          \fmfright{R}
          \fmftop{T}
          \fmf{fermion}{L,V,R}
          \fmf{phantom}{V,T}
          \fmffreeze
          \fmf{dbl_wiggly,right=1}{V,T,V}
          \fmfdot{V}
        \end{fmfgraph*}
      \end{fmffile}
    \end{gathered}
    \hspace{5pt} + \hspace{5pt}
    \begin{gathered}
      \begin{fmffile}{Loop_Fermion_4}
        \begin{fmfgraph*}(40,40)
          \fmfleft{L}
          \fmfright{R}
          \fmf{fermion,tension=2}{L,VL}
          \fmf{fermion,tension=2}{VR,R}
          \fmf{fermion}{VL,VR}
          \fmffreeze
          \fmf{dbl_wiggly,left=1}{VL,VR}
          \fmfdot{VL,VR}
        \end{fmfgraph*}
      \end{fmffile}
    \end{gathered}
    \\
    =& \cfrac{1}{8}\,\left[ \left(1 + 2\,\cfrac{m_\text{f}^2}{p^2}\right) \widehat{p} - 3\,m_\text{f}\right]\,\kappa^2\,i\,\pi^2\,A_0(m_\text{f}^2)\\
    & + \cfrac{1}{8}\,\left[ -\left( 1 - \cfrac{m_\text{f}^2}{p^2} + 2 \, \cfrac{m_\text{f}^4}{p^4} \right)\,p^2\,\widehat{p}  + m_\text{f} \left(p^2 + m_\text{f}^2\right) \right]\,\kappa^2\,i\,\pi^2\, B_0(p^2,0,m_\text{f}^2).
  \end{split}
\end{align}
Polarization operator for a Proca field:
\begin{align}
  \begin{split}
    i\,\Pi^{s=1,m_\text{v}\not =0} =& \hspace{20pt}
    \begin{gathered}
      \begin{fmffile}{Loop_Proca_3}
        \begin{fmfgraph*}(30,30)
          \fmfleft{L}
          \fmfright{R}
          \fmftop{T}
          \fmf{photon}{L,V,R}
          \fmf{phantom}{V,T}
          \fmffreeze
          \fmf{dbl_wiggly,right=1}{V,T,V}
          \fmfdot{V}
          \fmflabel{$\mu$}{L}
          \fmflabel{$\nu$}{R}
        \end{fmfgraph*}
      \end{fmffile}
    \end{gathered}
    \hspace{20pt} + \hspace{20pt}
    \begin{gathered}
      \begin{fmffile}{Loop_Proca_4}
        \begin{fmfgraph*}(30,30)
          \fmfleft{L}
          \fmfright{R}
          \fmf{photon,tension=2}{L,VL}
          \fmf{photon,tension=2}{VR,R}
          \fmf{photon,tension=.1}{VL,VR}
          \fmf{dbl_wiggly,left=1}{VL,VR}
          \fmfdot{VL,VR}
          \fmflabel{$\mu$}{L}
          \fmflabel{$\nu$}{R}
        \end{fmfgraph*}
      \end{fmffile}
    \end{gathered}
    \\
    =& \cfrac{1}{6}\,\Big\{ \left(p^2 - 8\, m_\text{v}^2\right) \,\kappa^2\,i\,\pi^2\,A_0(m_\text{v}^2) - \left(p^4 + p^2\,m_\text{v}^2 + m_\text{v}^4\right)\,\kappa^2\,i\,\pi^2\,B_0(p^2,0,m_\text{v}^2) \Big\}\theta_{\mu\nu}(p)\\
    & - \cfrac32\,m_\text{v}^4\,\kappa^2\,i\,\pi^2\,B_0(p^2,0,m_\text{v}^2) \, \omega_{\mu\nu}(p).
  \end{split}
\end{align}
Here the following definitions of gauge projectors are used:
\begin{align}
  \theta_{\mu\nu} (p) & = \eta_{\mu\nu} - \cfrac{p_\mu p_\nu}{p^2} \, , & \omega_{\mu\nu}(p) &= \cfrac{p_\mu p_\nu}{p^2}\,.
\end{align}
The polarization operator for a massless vector field reads:
\begin{align}
  \begin{split}
    \Pi^{s=1,m=0}_{\mu\nu} =& \hspace{20pt}
    \begin{gathered}
      \begin{fmffile}{Loop_Maxwell_5}
        \begin{fmfgraph*}(40,50)
          \fmfleft{L}
          \fmfright{R}
          \fmftop{T}
          \fmf{photon}{L,V,R}
          \fmf{phantom}{V,T}
          \fmffreeze
          \fmf{dbl_wiggly,right=1}{V,T,V}
          \fmfdot{V}
          \fmflabel{$\mu$}{L}
          \fmflabel{$\nu$}{R}
        \end{fmfgraph*}
      \end{fmffile}
    \end{gathered}
    \hspace{20pt} + \hspace{20pt}
    \begin{gathered}
      \begin{fmffile}{Loop_Maxwell_6}
        \begin{fmfgraph*}(40,40)
          \fmfleft{L}
          \fmfright{R}
          \fmf{photon,tension=2}{L,VL}
          \fmf{photon,tension=2}{VR,R}
          \fmf{photon,tension=.1}{VL,VR}
          \fmf{dbl_wiggly,left=1}{VL,VR}
          \fmfdot{VL,VR}
          \fmflabel{$\mu$}{L}
          \fmflabel{$\nu$}{R}
        \end{fmfgraph*}
      \end{fmffile}
    \end{gathered}
    \\
    =& -\cfrac{1}{6} \, \kappa^2\,i\,\pi^2\,B_0(p^2,0,0)\,p^4\,\theta_{\mu\nu}(p) - \cfrac{5}{2} \,\kappa^2\,i\,\pi^2\,B_0(p^2,0,0)\,p^4\,\omega_{\mu\nu}(p).
  \end{split}
\end{align}
Lastly, for the $SU(N)$ Yang-Mills theory there are polarization operators for gluons:
\begin{align}
  \begin{split}
    i\,\Pi_{\mu\nu}^\text{Gluon}=& \hspace{20pt}
    \begin{gathered}
      \begin{fmffile}{Loop_SUNYM_7}
        \begin{fmfgraph*}(40,50)
          \fmfleft{L}
          \fmfright{R}
          \fmftop{T}
          \fmf{gluon}{L,V,R}
          \fmf{phantom}{V,T}
          \fmffreeze
          \fmf{dbl_wiggly,right=1}{V,T,V}
          \fmfdot{V}
          \fmflabel{$\mu$}{L}
          \fmflabel{$\nu$}{R}
        \end{fmfgraph*}
      \end{fmffile}
    \end{gathered}
    \hspace{20pt} + \hspace{20pt}
    \begin{gathered}
      \begin{fmffile}{Loop_SUNYM_8}
        \begin{fmfgraph*}(40,40)
          \fmfleft{L}
          \fmfright{R}
          \fmf{gluon,tension=2}{L,VL}
          \fmf{gluon,tension=2}{VR,R}
          \fmf{gluon,tension=.1}{VL,VR}
          \fmf{dbl_wiggly,left=1}{VL,VR}
          \fmfdot{VL,VR}
          \fmflabel{$\mu$}{L}
          \fmflabel{$\nu$}{R}
        \end{fmfgraph*}
      \end{fmffile}
    \end{gathered}\\
    =&-\cfrac{1}{6}\, \kappa^2\,i\,\pi^2\,B_0(p^2,0,m^2)\,\delta^{ab}\, p^4 \,\left[ \theta_{\mu\nu}(p) + 15\, \omega_{\mu\nu}(p) \right];
  \end{split}
\end{align}
and polarization operators for quarks:
\begin{align}
  \begin{split}
    i\,\Pi_{\mu\nu}^\text{Quark} = & \hspace{10pt}
    \begin{gathered}
      \begin{fmffile}{Loop_SUNYM_9}
        \begin{fmfgraph*}(40,50)
          \fmfleft{L}
          \fmfright{R}
          \fmftop{T}
          \fmf{fermion}{L,V,R}
          \fmf{phantom}{V,T}
          \fmffreeze
          \fmf{dbl_wiggly,right=1}{V,T,V}
          \fmfdot{V}
        \end{fmfgraph*}
      \end{fmffile}
    \end{gathered}
    \hspace{5pt} + \hspace{5pt}
    \begin{gathered}
      \begin{fmffile}{Loop_SUNYM_10}
        \begin{fmfgraph*}(40,40)
          \fmfleft{L}
          \fmfright{R}
          \fmf{fermion}{L,VL}
          \fmf{fermion}{VR,R}
          \fmf{fermion}{VL,VR}
          \fmffreeze
          \fmf{dbl_wiggly,left=1}{VL,VR}
          \fmfdot{VL,VR}
        \end{fmfgraph*}
      \end{fmffile}
    \end{gathered}\\
    =& -\cfrac{1}{8}\, \kappa^2\,i\,\pi^2\,B_0(p^2,0,m^2)\, p^2\, \widehat{p}.
  \end{split}
\end{align}

\subsection{Example of a vertex operator}

Let us briefly consider another example of calculations that can be performed within FeynGrav. For the sake of illustration, we can address a one-loop scalar-graviton vertex function:
\begin{align}
  \nonumber \\ 
  i\,\Gamma_{\mu\nu}(k,p_1,p_2)= \hspace{20pt}
  \begin{gathered}
    \begin{fmffile}{Vertex}
      \begin{fmfgraph*}(40,40)
        \fmfleft{i}
        \fmfright{o1,o2}
        \fmf{dbl_wiggly}{i,v}
        \fmf{dashes}{o1,v}
        \fmf{dashes}{v,o2}
        \fmfv{decor.shape=circle,decor.filled=shaded,decor.size=20}{v}
        \fmflabel{$\mu\nu$}{i}
        \fmflabel{$p_1$}{o1}
        \fmflabel{$p_2$}{o2}
      \end{fmfgraph*}
    \end{fmffile}
  \end{gathered} \hspace{10pt}
  =
  \begin{gathered}
    \begin{fmffile}{Vertex_1}
      \begin{fmfgraph}(40,40)
        \fmfleft{i}
        \fmfright{o1,o2}
        \fmf{dbl_wiggly,tension=2}{i,v}
        \fmf{dashes,tension=0.7}{o1,v}
        \fmf{dashes,tension=0.7}{v,o2}
        \fmffreeze
        \fmf{phantom,tension=2}{o1,v1}
        \fmf{phantom,tension=2}{o2,v2}
        \fmf{phantom}{v1,v,v2}
        \fmffreeze
        \fmf{dbl_wiggly}{v1,v2}
        \fmfdot{v1,v2,v}
      \end{fmfgraph}
    \end{fmffile}
  \end{gathered}
  +
  \begin{gathered}
    \begin{fmffile}{Vertex_2}
      \begin{fmfgraph}(40,40)
        \fmfleft{i}
        \fmfright{o1,o2}
        \fmf{dbl_wiggly,tension=2}{i,v}
        \fmf{phantom,tension=0.7}{o1,v}
        \fmf{phantom,tension=0.7}{v,o2}
        \fmffreeze
        \fmf{dashes,tension=2}{o1,v1}
        \fmf{dashes,tension=2}{o2,v2}
        \fmf{dbl_wiggly}{v1,v,v2}
        \fmffreeze
        \fmf{dashes}{v1,v2}
        \fmfdot{v1,v2,v}
      \end{fmfgraph}
    \end{fmffile}
  \end{gathered}
  +
  \begin{gathered}
    \begin{fmffile}{Vertex_3}
      \begin{fmfgraph}(40,40)
        \fmfleft{i}
        \fmfright{o1,o2}
        \fmf{dbl_wiggly,tension=2}{i,v}
        \fmf{dashes,tension=0.7}{o1,v}
        \fmf{dashes,tension=0.7}{v,o2}
        \fmffreeze
        \fmf{phantom,tension=2}{o1,v1}
        \fmf{phantom}{v1,v}
        \fmfdot{v,v1}
        \fmffreeze
        \fmf{dbl_wiggly,right=0.7}{v1,v}
      \end{fmfgraph}
    \end{fmffile}
  \end{gathered}
  +
  \begin{gathered}
    \begin{fmffile}{Vertex_4}
      \begin{fmfgraph}(40,40)
        \fmfleft{i}
        \fmfright{o1,o2}
        \fmf{dbl_wiggly,tension=2}{i,v}
        \fmf{dashes,tension=0.7}{o1,v}
        \fmf{dashes,tension=0.7}{v,o2}
        \fmffreeze
        \fmf{phantom,tension=2}{o2,v1}
        \fmf{phantom}{v1,v}
        \fmfdot{v,v1}
        \fmffreeze
        \fmf{dbl_wiggly,left=0.7}{v1,v}
      \end{fmfgraph}
    \end{fmffile}
  \end{gathered}
  +
  \begin{gathered}
    \begin{fmffile}{Vertex_5}
      \begin{fmfgraph}(40,40)
        \fmfleft{i}
        \fmfright{o1,p,o2}
        \fmf{dbl_wiggly,tension=2}{i,v}
        \fmf{dashes,tension=0.5,left=0.5}{o1,v}
        \fmf{dashes,tension=0.5,right=0.5}{o2,v}
        \fmffreeze
        \fmf{phantom,tension=2}{p,p1}
        \fmf{dbl_wiggly,right=1}{v,p1,v}
        \fmfdot{v}
      \end{fmfgraph}
    \end{fmffile}
  \end{gathered} . \\ \nonumber
\end{align}
A detailed discussion of this function lies far beyond the scope of this paper and will be presented elsewhere. Here we will only consider a very specific limit of this function that was already studied in \cite{Donoghue:1994dn} (see also \cite{Burgess:2003jk,Vanhove:2021zel,Bjerrum-Bohr:2022blt} for a detailed discussion). We only consider the case when both scalars are placed on the mass shell $p_1^2=p_2^2=m^2$ and the graviton four-momentum only has spacial components $k^2 = - (\vec{k})^2$ and they are small. This setup allows one to recover the classical limit of the theory.

The example file``FeynGrav\_Examples.nb'' contains expressions calculating all the amplitude given above. Using FeynCalc tools we separate the Passarino-Veltman integrals and keep terms relevant for $\abs{\vec{k}}\to 0$ limit:
\begin{align}
  i\,\Gamma_{\mu\nu} (k,p_1,p_2) \to i\,\pi^2\,\kappa^3\,(p_1+p_2)_\mu (p_1+p_2)_\nu \left[ \cfrac{101}{96} \,\ln k^2 + \cfrac{\pi^2}{32} \, \cfrac{m}{k} \right] .
\end{align}
This expression is in agreement with the previous studies in \cite{Donoghue:1994dn,Vanhove:2021zel,Bjerrum-Bohr:2022blt}, where the leading-order contributions are non-analytic functions that correspond to power-law corrections to the Newtonian potential.

\subsection{Example of a tree-level scattering amplitude}

Lastly, we want to briefly touch upon the implementation of FeynGrav for scattering amplitudes. In full analogy with the previous case, a more detailed discussion of scattering amplitudes lies far beyond the scale of this paper. Because of this, we will only consider a single tree-level scattering amplitude for two scalars of different masses.

\begin{align}
  i \, \mathcal{M} (p_1,p_2,p_3,p_4,m_1,m_2) = \hspace{40pt}
  \begin{gathered}
    \begin{fmffile}{Scalar_Tree_Scattering}
      \begin{fmfgraph*}(50,50)
        \fmftop{t1,t2}
        \fmfbottom{b1,b2}
        \fmf{dashes}{t1,v1,b1}
        \fmf{dashes}{t2,v2,b2}
        \fmf{dbl_wiggly}{v1,v2}
        \fmfdot{v1,v2}
        \fmflabel{$p_1,m_1$}{b1}
        \fmflabel{$p_2,m_2$}{b2}
        \fmflabel{$p_3,m_1$}{t1}
        \fmflabel{$p_4,m_2$}{t2}
      \end{fmfgraph*}
    \end{fmffile}
    \hspace{40pt} .
  \end{gathered}
\end{align}
It is more convenient to express this amplitude in terms of the Mandelstam variables \cite{Mandelstam:1958xc}:
\begin{align}
  s &= (p_1+p_2)^2 \,, & t &= (p_1 + p_3)^2 \,, & u &= (p_1+p_4)^2 .
\end{align}
The scattering amplitude reads:
\begin{align}
  i\,\mathcal{M} = -i\, \cfrac{ \kappa^2}{4\,t} ~ \Big( u^2 + t\, u - (t +2\,u)(m_1^2 + m_2^2) + m_1^4 + m_2^4 \Big).
\end{align}
In full analogy with the previous case, it is convenient to consider a quasi-static limit:
\begin{align}
  s &= (m_1+m_2)^2\,, & t &= -\left(\vec{p}\right)^2 \to 0 \,.
\end{align}
That limit recovers the part amplitude leading in the weak interaction limit which reads
\begin{align}
  \mathcal{M} \sim \cfrac{\kappa^2}{2} \, \cfrac{m_1^2\,m_2^2}{p^2}\,.
\end{align}
In full agreement with \cite{Donoghue:1994dn,Burgess:2003jk,Vanhove:2021zel,Bjerrum-Bohr:2022blt} the recovered contribution corresponds to the leading-order term in the Newtonian potential.

\section{Conclusions}\label{section_conclusions}

In this paper, we present the latest developments of FeynGrav, which offers a simple and efficient way to derive Feynman's rules for gravity. Building on our previous work in \cite{Latosh:2022ydd}, where we derived the Feynman rules for gravitational interaction with massless matter, we extend the formalism to cover matter with arbitrary mass. We also revisit the implementation of the Faddeev-Popov prescription within the formalism and derive the corresponding rules for the Faddeev-Popov ghosts present in the theory of a single massless vector field. Additionally, we implement the formalism to the $SU(N)$ Yang-Mills model and obtain all the required interaction rules. These interaction rules are sufficient for calculating gravitational corrections to standard model processes, which opens up new opportunities to search for relevant gravitational effects within the standard model.

The explicit examples of tree and loop-level calculations performed with FeynGrav demonstrate the usefulness of the presented rules, and the potential for further applications of FeynGrav for scattering amplitudes is promising. The contemporary methods of scattering amplitude calculations are well-developed for on-shell amplitudes \cite{Elvang:2013cua,Cheung:2017pzi,Travaglini:2022uwo}. FeynGrav provides a way to calculate off-shell scattering amplitudes, which is an important step toward studying higher-order effects in gravitational interactions.

Future developments of FeynGrav will focus on several directions. First, we plan to implement non-minimal interactions, particularly non-minimal interactions with scalar fields \cite{Horndeski:1974wa,Horndeski:1976gi,Kobayashi:2011nu}. This will allow us to study their influence on quantum gravity behavior. Secondly, we aim to improve the performance of the package, as quantum gravitational calculations are notoriously complicated due to a large number of terms and Lorentz indices involved. We plan to explore techniques such as parallel computations to increase FeynGrav's performance. Lastly, we intend to extend the formalism to supersymmetric models, which will provide another effective tool to operate with supergravity scattering amplitudes.

\section*{Acknowledgment}
The work was supported by the Foundation for the Advancement of Theoretical Physics and Mathematics “BASIS”.

\bibliographystyle{unsrturl}
\bibliography{FG2.bib}

\newpage
\appendix

\section{FeynGrav commands}\label{Command_list}

\begin{table}[h!]
  \begin{center}
    \begin{tabular}{c|c}
      Diagram & Command \\ \hline
      \begin{minipage}{.2\textwidth}
        \begin{align*}
          \\
          \begin{fmffile}{IS0}
            \begin{fmfgraph*}(30,10)
              \fmfleft{L}
              \fmfright{R}
              \fmf{dashes}{L,R}
            \end{fmfgraph*}
          \end{fmffile}
          \\
        \end{align*}
      \end{minipage}
      &
      ScalarPropagator$[p,m]$ \\ \hline
      \begin{minipage}{.2\textwidth}
        \begin{align*}
          \\
          \begin{fmffile}{IS1}
            \begin{fmfgraph*}(30,30)
              \fmfleft{L}
              \fmfright{R1,R2}
              \fmf{dbl_wiggly}{L,V}
              \fmf{dashes}{R1,V,R2}
              \fmfdot{V}
              \fmflabel{$\mu\nu$}{L}
              \fmflabel{$p_1$}{R1}
              \fmflabel{$p_2$}{R2}
            \end{fmfgraph*}
          \end{fmffile}
          \\
        \end{align*}
      \end{minipage}
      &
      GravitonScalarVertex$[\{\mu,\nu\},p_1,p_2,m]$\\ \hline
      \begin{minipage}{.2\textwidth}
        \begin{align*}
          \\
          \begin{fmffile}{IS2}
            \begin{fmfgraph*}(30,30)
              \fmfleft{L1,L2}
              \fmfright{R1,R2}
              \fmf{dbl_wiggly}{L1,V,L2}
              \fmf{dashes}{R1,V,R2}
              \fmfdot{V}
              \fmflabel{$\mu\nu$}{L1}
              \fmflabel{$\alpha\beta$}{L2}
              \fmflabel{$p_1$}{R1}
              \fmflabel{$p_2$}{R2}
            \end{fmfgraph*}
          \end{fmffile}
          \\
        \end{align*}
      \end{minipage}
      &
      GravitonScalarVertex$[\{\mu,\nu,\alpha,\beta\},p_1,p_2,m]$ \\ \hline
      \begin{minipage}{.2\textwidth}
        \begin{align*}
          \\
          \begin{fmffile}{IS3}
            \begin{fmfgraph*}(30,30)
              \fmfleft{L}
              \fmfright{R1,R2,R3}
              \fmf{dbl_wiggly,tension=2,label=$\lambda_3$}{L,V}
              \fmf{dashes}{R1,V}
              \fmf{dashes}{R2,V}
              \fmf{dashes}{R3,V}
              \fmfdot{V}
              \fmflabel{$\mu\nu$}{L}
              \fmflabel{$p_1$}{R1}
              \fmflabel{$p_2$}{R2}
              \fmflabel{$p_3$}{R3}
            \end{fmfgraph*}
          \end{fmffile}
          \\
        \end{align*}
      \end{minipage}
      &
      GravitonScalarPotentialVertex$[\{\mu,\nu\},\lambda_3]$ \\ \hline
      \begin{minipage}{.2\textwidth}
        \begin{align*}
          \\
          \begin{fmffile}{IS4}
            \begin{fmfgraph*}(30,30)
              \fmfleft{L}
              \fmfright{R1,R2,R3,R4}
              \fmf{dbl_wiggly,tension=3,label=$\lambda_4$}{L,V}
              \fmf{dashes}{R1,V}
              \fmf{dashes}{R2,V}
              \fmf{dashes}{R3,V}
              \fmf{dashes}{R4,V}
              \fmfdot{V}
              \fmflabel{$\mu\nu$}{L}
              \fmflabel{$p_1$}{R1}
              \fmflabel{$p_2$}{R2}
              \fmflabel{$p_3$}{R3}
              \fmflabel{$p_4$}{R4}
            \end{fmfgraph*}
          \end{fmffile}
          \\
        \end{align*}
      \end{minipage}
      &
      GravitonScalarPotentialVertex$[\{\mu,\nu\},\lambda_4]$ \\ \hline
      \begin{minipage}{.2\textwidth}
        \begin{align*}
          \\
          \begin{fmffile}{IS5}
            \begin{fmfgraph*}(30,30)
              \fmfleft{L1,L2}
              \fmfright{R1,R2,R3,R4}
              \fmf{dbl_wiggly,tension=2,label=$\lambda_4$}{L1,V}
              \fmf{dbl_wiggly,tension=2}{L2,V}
              \fmf{dashes}{R1,V}
              \fmf{dashes}{R2,V}
              \fmf{dashes}{R3,V}
              \fmf{dashes}{R4,V}
              \fmfdot{V}
              \fmflabel{$\mu\nu$}{L1}
              \fmflabel{$\alpha\beta$}{L2}
              \fmflabel{$p_1$}{R1}
              \fmflabel{$p_2$}{R2}
              \fmflabel{$p_3$}{R3}
              \fmflabel{$p_4$}{R4}
            \end{fmfgraph*}
          \end{fmffile}
          \\
        \end{align*}
      \end{minipage}
      &
      GravitonScalarPotentialVertex$[\{\mu,\nu,\alpha,\beta\},\lambda_4]$ \\ \hline 
      \begin{minipage}{.2\textwidth}
        \begin{align*}
          \\
          \begin{fmffile}{IF1}
            \begin{fmfgraph*}(30,30)
              \fmfleft{L}
              \fmfright{R1,R2}
              \fmf{dbl_wiggly}{L,V}
              \fmf{fermion}{R1,V,R2}
              \fmfdot{V}
              \fmflabel{$\mu\nu$}{L}
              \fmflabel{$p_1$}{R1}
              \fmflabel{$p_2$}{R2}
            \end{fmfgraph*}
          \end{fmffile}
          \\
        \end{align*}
      \end{minipage}
      &
      GravitonFermionVertex$[\{\mu,\nu\},p_1,p_2,m]$\\ \hline
      \begin{minipage}{.2\textwidth}
        \begin{align*}
          \\
          \begin{fmffile}{IF2}
            \begin{fmfgraph*}(30,30)
              \fmfleft{L1,L2}
              \fmfright{R1,R2}
              \fmf{dbl_wiggly}{L1,V,L2}
              \fmf{fermion}{R1,V,R2}
              \fmfdot{V}
              \fmflabel{$\mu\nu$}{L1}
              \fmflabel{$\alpha\beta$}{L2}
              \fmflabel{$p_1$}{R1}
              \fmflabel{$p_2$}{R2}
            \end{fmfgraph*}
          \end{fmffile}
          \\
        \end{align*}
      \end{minipage}
      &
      GravitonFermionVertex$[\{\mu,\nu,\alpha,\beta\},p_1,p_2,m]$ \\ \hline
    \end{tabular}
  \end{center}
\end{table}

\newpage

\begin{table}[h!]
  \begin{center}
    \begin{tabular}{c|c}
      Diagram & Command \\ \hline
      \begin{minipage}{.2\textwidth}
        \begin{align*}
          \\
          \begin{fmffile}{IV0}
            \begin{fmfgraph*}(30,10)
              \fmfleft{L}
              \fmfright{R}
              \fmf{photon,label=$m\not=0$}{L,R}
              \fmflabel{$\mu$}{L}
              \fmflabel{$\nu$}{R}
            \end{fmfgraph*}
          \end{fmffile}
          \\
        \end{align*}
      \end{minipage}
      &
      ProcaPropagator$[\mu,\nu,p,m]$ \\ \hline
      \begin{minipage}{.2\textwidth}
        \begin{align*}
          \\
          \begin{fmffile}{IV1}
            \begin{fmfgraph*}(30,30)
              \fmfleft{L}
              \fmfright{R1,R2}
              \fmf{dbl_wiggly}{L,V}
              \fmf{photon}{R1,V,R2}
              \fmfdot{V}
              \fmflabel{$\rho\sigma,k$}{L}
              \fmflabel{$\mu,p_1$}{R1}
              \fmflabel{$\nu,p_2$}{R2}
            \end{fmfgraph*}
          \end{fmffile}
          \\
        \end{align*}
      \end{minipage}
      &
      GravitonVectorVertex$[\{\rho,\sigma,k\},\mu,p_1,\nu,p_2]$\\ \hline
      \begin{minipage}{.2\textwidth}
        \begin{align*}
          \\
          \begin{fmffile}{IV2}
            \begin{fmfgraph*}(30,30)
              \fmfleft{L1,L2}
              \fmfright{R1,R2}
              \fmf{dbl_wiggly}{L1,V,L2}
              \fmf{photon}{R1,V,R2}
              \fmfdot{V}
              \fmflabel{$\rho_1\sigma_1,k_1$}{L1}
              \fmflabel{$\rho_2\sigma_2,k_2$}{L2}
              \fmflabel{$\mu,p_1$}{R1}
              \fmflabel{$\nu,p_2$}{R2}
            \end{fmfgraph*}
          \end{fmffile}
          \\
        \end{align*}
      \end{minipage}
      &
      GravitonVectorVertex$[\{\rho_1,\sigma_1,k_1,\rho_2\sigma_2,k_2\},\mu,p_1,\nu,p_2]$ \\ \hline
      \begin{minipage}{.2\textwidth}
        \begin{align*}
          \\
          \begin{fmffile}{IV3}
            \begin{fmfgraph*}(30,30)
              \fmfleft{L}
              \fmfright{R1,R2}
              \fmf{dbl_wiggly}{L,V}
              \fmf{dots}{R1,V}
              \fmf{dots}{R2,V}
              \fmfdot{V}
              \fmflabel{$\rho\sigma$}{L}
              \fmflabel{$p_1$}{R1}
              \fmflabel{$p_2$}{R2}
            \end{fmfgraph*}
          \end{fmffile}
          \\
        \end{align*}
      \end{minipage}
      &
      GravitonVectorGhostVertex$[\{\rho,\sigma\},p_1,p_2]$ \\ \hline
      \begin{minipage}{.2\textwidth}
        \begin{align*}
          \\
          \begin{fmffile}{IYM1}
            \begin{fmfgraph*}(30,30)
              \fmfleft{L}
              \fmfright{R1,R2}
              \fmf{dbl_wiggly,tension=2}{L,V}
              \fmf{gluon}{R1,V}
              \fmf{gluon}{R2,V}
              \fmfdot{V}
              \fmflabel{$\rho\sigma,k$}{L}
              \fmflabel{$p_1,\lambda_1,a_1$}{R1}
              \fmflabel{$p_2,\lambda_2,a_2$}{R2}
            \end{fmfgraph*}
          \end{fmffile}
          \\
        \end{align*}
      \end{minipage}
      &
      GravitonGluonVertex$[\{\rho,\sigma,k\},p_1,\lambda_1,a_1,p_2,\lambda_2,a_2]$ \\ \hline
      \begin{minipage}{.2\textwidth}
        \begin{align*}
          \\
          \begin{fmffile}{IYM2}
            \begin{fmfgraph*}(30,30)
              \fmfleft{L}
              \fmfright{R1,R2,R3}
              \fmf{dbl_wiggly,tension=2}{L,V}
              \fmf{gluon,tension=.5}{R1,V}
              \fmf{gluon,tension=.5}{R2,V}
              \fmf{gluon,tension=.5}{R3,V}
              \fmfdot{V}
              \fmflabel{$\rho\sigma,k$}{L}
              \fmflabel{$p_1,\lambda_1,a_1$}{R1}
              \fmflabel{$p_2,\lambda_2,a_2$}{R2}
              \fmflabel{$p_3,\lambda_3,a_3$}{R3}
            \end{fmfgraph*}
          \end{fmffile}
          \\
        \end{align*}
      \end{minipage}
      &
      GravitonGluonVertex$[\{\rho,\sigma,k\},p_1,\lambda_1,a_1,p_2,\lambda_2,a_2,p_3,\lambda_3,a_3]$ \\ \hline
      \begin{minipage}{.3\textwidth}
        \begin{align*}
          \\
          \begin{fmffile}{IYM3}
            \begin{fmfgraph*}(40,40)
              \fmfleft{L}
              \fmfright{R1,R2,R3,R4}
              \fmf{dbl_wiggly,tension=2}{L,V}
              \fmf{gluon,tension=.4}{R1,V}
              \fmf{gluon,tension=.4}{R2,V}
              \fmf{gluon,tension=.4}{R3,V}
              \fmf{gluon,tension=.4}{R4,V}
              \fmfdot{V}
              \fmflabel{$\rho\sigma,k$}{L}
              \fmflabel{$p_1,\lambda_1,a_1$}{R1}
              \fmflabel{$p_2,\lambda_2,a_2$}{R2}
              \fmflabel{$p_3,\lambda_3,a_3$}{R3}
              \fmflabel{$p_4,\lambda_4,a_4$}{R4}
            \end{fmfgraph*}
          \end{fmffile}
          \\
        \end{align*}
      \end{minipage}
      &
      GravitonGluonVertex$[\{\rho,\sigma,k\},p_1,\lambda_1,a_1,p_2,\lambda_2,a_2,p_3,\lambda_3,a_3,p_4,\lambda_4,a_4]$ \\ \hline
      \begin{minipage}{.2\textwidth}
        \begin{align*}
          \\
          \begin{fmffile}{IYM4}
            \begin{fmfgraph*}(30,40)
              \fmfleft{L}
              \fmfright{R1,R2,R3}
              \fmf{dbl_wiggly,tension=2}{L,V}
              \fmf{fermion,tension=.5}{R1,V}
              \fmf{gluon,tension=.5}{R2,V}
              \fmf{fermion,tension=.5}{V,R3}
              \fmfdot{V}
              \fmflabel{$\rho\sigma$}{L}
              \fmflabel{$\lambda,a$}{R2}
            \end{fmfgraph*}
          \end{fmffile}
          \\
        \end{align*}
      \end{minipage}
      &
      GravitonQuarkGluonVertex$[\{\rho,\sigma\},\lambda,a]$ \\ \hline
    \end{tabular}
  \end{center}
\end{table}

\newpage

\begin{table}[h!]
  \begin{center}
    \begin{tabular}{c|c}
      Diagram & Command \\ \hline
      \begin{minipage}{.2\textwidth}
        \begin{align*}
          \\
          \begin{fmffile}{IYM5}
            \begin{fmfgraph*}(30,40)
              \fmfleft{L}
              \fmfright{R1,R2}
              \fmf{dbl_wiggly,tension=2}{L,V}
              \fmf{dots_arrow,tension=.5}{R1,V}
              \fmf{dots_arrow,tension=.5}{V,R2}
              \fmfdot{V}
              \fmflabel{$\rho\sigma$}{L}
              \fmflabel{$p_2, a_2$}{R2}
              \fmflabel{$p_1, a_1$}{R1}
            \end{fmfgraph*}
          \end{fmffile}
          \\
        \end{align*}
      \end{minipage}
      &
      GravitonYMGhostVertex$[\{\rho,\sigma\},p_1,a_1,p_2,a_2]$ \\ \hline
      \begin{minipage}{.3\textwidth}
        \begin{align*}
          \\
          \begin{fmffile}{IYM6}
            \begin{fmfgraph*}(30,40)
              \fmfleft{L}
              \fmfright{R1,R2,R3}
              \fmf{dbl_wiggly,tension=2}{L,V}
              \fmf{dots_arrow,tension=.5}{R1,V}
              \fmf{gluon,tension=.5}{R2,V}
              \fmf{dots_arrow,tension=.5}{V,R3}
              \fmfdot{V}
              \fmflabel{$\rho\sigma$}{L}
              \fmflabel{$p_2,a_2$}{R1}
              \fmflabel{$\mu_1,p_1,a_1$}{R2}
              \fmflabel{$p_3,a_3$}{R3}
            \end{fmfgraph*}
          \end{fmffile}
          \\
        \end{align*}
      \end{minipage}
      &
      GravitonGluonGhostVertex$[\{\rho,\sigma\},p_1,\mu_1,a_1,p_2,\mu_2,a_2,p_3,\mu_3,a_3]$ \\ \hline
      \begin{minipage}{.2\textwidth}
        \begin{align*}
          \\
          \begin{fmffile}{IG0}
            \begin{fmfgraph*}(30,40)
              \fmfleft{L}
              \fmfright{R}
              \fmf{dbl_wiggly}{L,R}
              \fmflabel{$\mu\nu$}{L}
              \fmflabel{$\alpha\beta$}{R}
            \end{fmfgraph*}
          \end{fmffile}
          \\
        \end{align*}
      \end{minipage}
      &
      GravitonPropagator$[\mu,\nu,\alpha,\beta,p]$ \\ \hline
      \begin{minipage}{.2\textwidth}
        \begin{align*}
          \\
          \begin{fmffile}{IG1}
            \begin{fmfgraph*}(30,30)
              \fmfleft{L}
              \fmfright{R1,R2}
              \fmf{dbl_wiggly}{L,V}
              \fmf{dbl_wiggly}{R1,V}
              \fmf{dbl_wiggly}{R2,V}
              \fmflabel{$\mu_1\nu_1,p_1$}{L}
              \fmflabel{$\mu_2\nu_2,p_2$}{R1}
              \fmflabel{$\mu_3\nu_3,p_3$}{R2}
              \fmfdot{V}
            \end{fmfgraph*}
          \end{fmffile}
          \\
        \end{align*}
      \end{minipage}
      &
      GravitonVertex$[\mu_1,\nu_1,p_1,\mu_2,\nu_2,p_2,\mu_3,\nu_3,p_3]$ \\ \hline
      \begin{minipage}{.2\textwidth}
        \begin{align*}
          \\
          \begin{fmffile}{IG2}
            \begin{fmfgraph*}(30,30)
              \fmfleft{L1,L2}
              \fmfright{R1,R2}
              \fmf{dbl_wiggly}{L1,V}
              \fmf{dbl_wiggly}{R1,V}
              \fmf{dbl_wiggly}{R2,V}
              \fmf{dbl_wiggly}{L2,V}
              \fmflabel{$\mu_1\nu_1,p_1$}{L1}
              \fmflabel{$\mu_2\nu_2,p_2$}{R1}
              \fmflabel{$\mu_3\nu_3,p_3$}{R2}
              \fmflabel{$\mu_4\nu_4,p_4$}{L2}
              \fmfdot{V}
            \end{fmfgraph*}
          \end{fmffile}
          \\
        \end{align*}
      \end{minipage}
      &
      GravitonVertex$[\mu_1,\nu_1,p_1,\mu_2,\nu_2,p_2,\mu_3,\nu_3,p_3,\mu_4,\nu_4,p_4]$ \\ \hline
      \begin{minipage}{.2\textwidth}
        \begin{align*}
          \\
          \begin{fmffile}{IG3}
            \begin{fmfgraph*}(30,30)
              \fmfleft{L}
              \fmfright{R1,R2}
              \fmf{dbl_wiggly}{L,V}
              \fmf{dots_arrow}{R1,V}
              \fmf{dots_arrow}{V,R2}
              \fmflabel{$\mu\nu,k$}{L}
              \fmflabel{$\lambda_1,p_1$}{R1}
              \fmflabel{$\lambda_2,p_2$}{R2}
              \fmfdot{V}
            \end{fmfgraph*}
          \end{fmffile}
          \\
        \end{align*}
      \end{minipage}
      &
      GravitonGhostVertex$[\{\mu,\nu,k\},\lambda_1,p_1,\lambda_2,p_2]$ \\ \hline
    \end{tabular}
  \end{center}
\end{table}

The package contains the following custom symbols used in calculations.
\begin{table}[h!]
  \begin{center}
    \begin{tabular}{c|c|c}
      Constant & Command & Comment \\ \hline & & \\
      $\kappa$ & \textbackslash[Kappa] &
      The gravitational coupling \eqref{the_gravitational_coupling_definition} \\ & \\ \hline & & \\
      $\varepsilon_\text{Vector}$ & GaugeFixingEpsilonVector &
      Gauge fixing parameter for a single vector \eqref{gauge_fixing_vector}   \\ & \\ \hline  & & \\
      $\varepsilon_\text{SU(N)YM}$ & GaugeFixingEpsilonSUNYM &
      Gauge fixing parameter for $SU(N)$ Yang-Mills \\ & \\ \hline & & \\
      $\varepsilon_\text{Gravity}$ & GaugeFixingEpsilon &
      Gauge fixing parameter for gravity \eqref{Hilbert_gauge_fixed} \\ & \\ \hline
    \end{tabular}
  \end{center}
\end{table}

\end{document}